\documentclass[english]{article}
\usepackage[T1]{fontenc}
\usepackage[latin9]{inputenc}
\usepackage{geometry}
\usepackage{float}
\usepackage{mathtools}
\usepackage{amsmath}
\usepackage{graphicx}
\PassOptionsToPackage{normalem}{ulem}
\usepackage{ulem}

\makeatletter
\newcommand{\lyxaddress}[1]{
	\par {\center #1
	\vspace{1.4em}
	\noindent\par}
}

\@ifundefined{date}{}{\date{}}
\RequirePackage{colortbl, tabularx}
\@ifundefined{comment}{}
  {
   
  }%

\makeatother

\usepackage{babel}
\begin{document}
\global\long\def\l{\lambda}%
\global\long\def\ints{\mathbb{Z}}%
\global\long\def\nat{\mathbb{N}}%
\global\long\def\re{\mathbb{R}}%
\global\long\def\com{\mathbb{C}}%
\global\long\def\dff{\triangleq}%
\global\long\def\df{\coloneqq}%
\global\long\def\del{\nabla}%
\global\long\def\cross{\times}%
\global\long\def\der#1#2{\frac{d#1}{d#2}}%
\global\long\def\bra#1{\left\langle #1\right|}%
\global\long\def\ket#1{\left|#1\right\rangle }%
\global\long\def\braket#1#2{\left\langle #1|#2\right\rangle }%
\global\long\def\ketbra#1#2{\left|#1\right\rangle \left\langle #2\right|}%
\global\long\def\paulix{\begin{pmatrix}0  &  1\\
 1  &  0 
\end{pmatrix}}%
\global\long\def\pauliy{\begin{pmatrix}0  &  -i\\
 i  &  0 
\end{pmatrix}}%
\global\long\def\pauliz{\begin{pmatrix}1  &  0\\
 0  &  -1 
\end{pmatrix}}%
\global\long\def\sinc{\mbox{sinc}}%
\global\long\def\ft{\mathcal{F}}%
\global\long\def\dg{\dagger}%
\global\long\def\bs#1{\boldsymbol{#1}}%
\global\long\def\norm#1{\left\Vert #1\right\Vert }%
\global\long\def\H{\mathcal{H}}%
\global\long\def\tens{\varotimes}%
\global\long\def\rationals{\mathbb{Q}}%
 
\global\long\def\tri{\triangle}%
\global\long\def\lap{\triangle}%
\global\long\def\e{\varepsilon}%
\global\long\def\broket#1#2#3{\bra{#1}#2\ket{#3}}%
\global\long\def\dv{\del\cdot}%
\global\long\def\eps{\epsilon}%
\global\long\def\rot{\vec{\del}\cross}%
\global\long\def\pd#1#2{\frac{\partial#1}{\partial#2}}%
\global\long\def\L{\mathcal{L}}%
\global\long\def\inf{\infty}%
\global\long\def\d{\delta}%
\global\long\def\I{\mathbb{I}}%
\global\long\def\D{\Delta}%
\global\long\def\r{\rho}%
\global\long\def\hb{\hbar}%
\global\long\def\s{\sigma}%
\global\long\def\t{\tau}%
\global\long\def\O{\Omega}%
\global\long\def\a{\alpha}%
\global\long\def\b{\beta}%
\global\long\def\th{\theta}%
\global\long\def\l{\lambda}%

\global\long\def\Z{\mathcal{Z}}%
\global\long\def\z{\zeta}%
\global\long\def\ord#1{\mathcal{O}\left(#1\right)}%
\global\long\def\ua{\uparrow}%
\global\long\def\da{\downarrow}%
 
\global\long\def\co#1{\left[#1\right)}%
\global\long\def\oc#1{\left(#1\right]}%
\global\long\def\tr{\mbox{tr}}%
\global\long\def\o{\omega}%
\global\long\def\nab{\del}%
\global\long\def\p{\psi}%
\global\long\def\pro{\propto}%
\global\long\def\vf{\varphi}%
\global\long\def\f{\phi}%
\global\long\def\mark#1#2{\underset{#2}{\underbrace{#1}}}%
\global\long\def\markup#1#2{\overset{#2}{\overbrace{#1}}}%
\global\long\def\ra{\rightarrow}%
\global\long\def\cd{\cdot}%
\global\long\def\v#1{\vec{#1}}%
\global\long\def\fd#1#2{\frac{\d#1}{\d#2}}%
\global\long\def\P{\Psi}%
\global\long\def\dem{\overset{\mbox{!}}{=}}%
\global\long\def\Lam{\Lambda}%
 
\global\long\def\m{\mu}%
\global\long\def\n{\nu}%

\global\long\def\ul#1{\underline{#1}}%
\global\long\def\at#1#2{\biggl|_{#1}^{#2}}%
\global\long\def\lra{\leftrightarrow}%
\global\long\def\var{\mbox{var}}%
\global\long\def\E{\mathcal{E}}%
\global\long\def\Op#1#2#3#4#5{#1_{#4#5}^{#2#3}}%
\global\long\def\up#1#2{\overset{#2}{#1}}%
\global\long\def\down#1#2{\underset{#2}{#1}}%
\global\long\def\lb{\biggl[}%
\global\long\def\rb{\biggl]}%
\global\long\def\RG{\mathfrak{R}_{b}}%
\global\long\def\g{\gamma}%
\global\long\def\Ra{\Rightarrow}%
\global\long\def\x{\xi}%
\global\long\def\c{\chi}%
\global\long\def\res{\mbox{Res}}%
\global\long\def\dif{\mathbf{d}}%
\global\long\def\dd{\mathbf{d}}%
\global\long\def\grad{\del}%
\global\long\def\floor#1{}%
\global\long\def\ceil#1{}%

\global\long\def\mat#1#2#3#4{\left(\begin{array}{cc}
#1 & #2\\
#3 & #4
\end{array}\right)}%
\global\long\def\col#1#2{\left(\begin{array}{c}
#1\\
#2
\end{array}\right)}%
\global\long\def\sl#1{\cancel{#1}}%
\global\long\def\row#1#2{\left(\begin{array}{cc}
#1 & ,#2\end{array}\right)}%
\global\long\def\roww#1#2#3{\left(\begin{array}{ccc}
#1 & ,#2 & ,#3\end{array}\right)}%
\global\long\def\rowww#1#2#3#4{\left(\begin{array}{cccc}
#1 & ,#2 & ,#3 & ,#4\end{array}\right)}%
\global\long\def\matt#1#2#3#4#5#6#7#8#9{\left(\begin{array}{ccc}
#1 & #2 & #3\\
#4 & #5 & #6\\
#7 & #8 & #9
\end{array}\right)}%
\global\long\def\su{\uparrow}%
\global\long\def\sd{\downarrow}%
\global\long\def\coll#1#2#3{\left(\begin{array}{c}
#1\\
#2\\
#3
\end{array}\right)}%
\global\long\def\h#1{\hat{#1}}%
\global\long\def\colll#1#2#3#4{\left(\begin{array}{c}
#1\\
#2\\
#3\\
#4
\end{array}\right)}%
\global\long\def\check{\checked}%
\global\long\def\v#1{\vec{#1}}%
\global\long\def\S{\Sigma}%
\global\long\def\F{\Phi}%
\global\long\def\M{\mathcal{M}}%
\global\long\def\G{\Gamma}%
\global\long\def\im{\mbox{Im}}%
\global\long\def\til#1{\tilde{#1}}%
\global\long\def\kb{k_{B}}%
\global\long\def\k{\kappa}%
\global\long\def\ph{\phi}%
\global\long\def\el{\ell}%
\global\long\def\en{\mathcal{N}}%
\global\long\def\asy{\cong}%
\global\long\def\sbl{\biggl[}%
\global\long\def\sbr{\biggl]}%
\global\long\def\cbl{\biggl\{}%
\global\long\def\cbr{\biggl\}}%
\global\long\def\hg#1#2{\mbox{ }_{#1}F_{#2}}%
\global\long\def\J{\mathcal{J}}%
\global\long\def\diag#1{\mbox{diag}\left[#1\right]}%
\global\long\def\sign#1{\mbox{sgn}\left[#1\right]}%
\global\long\def\T{\th}%
\global\long\def\rp{\reals^{+}}%

\title{Local resetting with geometric confinement}
\author{Asaf Miron}
\maketitle

\lyxaddress{Independent Researcher, Haifa 3276605, Israel}

\begin{abstract}
``Local resetting'' was recently introduced to describe stochastic
resetting in interacting systems where particles independently try
to reset to a common ``origin''. Our understanding of such systems,
where the resetting process is itself affected by interactions, is
still very limited. One ubiquitous constraint that is often imposed
on the dynamics of interacting particles is geometric confinement,
e.g. restricting rigid spherical particles to a channel so narrow
that overtaking becomes difficult. We here explore the interplay between
local resetting and geometric confinement in a system consisting of
two species of diffusive particles: ``bath'' particles, and ``tracers''
which undergo local resetting. Mean-field analysis and numerical simulations
show that the resetting tracers, whose stationary density profile
exhibits a typical ``tent-like'' shape, imprint this shape onto
the bath density profile. Upon varying the ratio of the degree of
geometric confinement over particle diffusivity, the system is found
to transition between two states. In one tracers expel bath particles
away from the origin, while in the other they ensnare them instead.
Between these two states, we find a special case where the mean field
approximation becomes exact.
\end{abstract}

\section{Introduction}

Stochastic resetting is a non-equilibrium process that is typically
used to model dynamical observables that are ``instantaneously''
\footnote{i.e. much faster than any other relevant time-scale in the problem.}
returned to their initial state some random time after having started
their temporal evolution \cite{Evans_2011,evans2013optimal,Majumdar_2014,Touchette_2015,bhat2016stochastic,Shlomi_2017,Chechkin_2018,tal2020experimental,evans2020stochastic}.
With applications ranging from search and optimization algorithms
\cite{Montanari_2002,benichou2011intermittent,Belan_2018,Bruyne_2020},
through chemical reactions \cite{reuveni2014role,Shlomi_2015,robin2018single},
to processes inside biological cells \cite{Eliazar_2007,Rold_2016,lisica2016mechanisms},
the study of stochastic resetting is attracting broad interest from
a diverse scientific audience.

Stochastic resetting has already been vigorously explored under a
variety of settings. These efforts have been greatly aided by the
``renewal'' framework, an analytical tool that allows deriving exact
results in situations where stochastic resetting acts ``globally''
- when stochastic resetting is either applied to a single degree of
freedom, or when it is simultaneously applied to multiple degrees
of freedom \cite{falcao2017interacting,evans2018run,Bodrova_2019,Kundu_2019,Karthika_2020}.
Perhaps due to its effectiveness and applicability to a wide range
of systems, far less attention has been directed towards scenarios
where stochastic resetting acts ``locally'' and the resetting process
is itself affected by interactions, i.e. when different interacting
degrees of freedom reset \textbf{independently} of one-another. Yet
in real complex systems whose dynamics are modeled by stochastic resetting,
there is no a-priori reason to assume interactions could not dramatically
affect resetting. Exploring the interplay between locally-acting stochastic
resetting and various forms of interaction is thus a natural step
forward, and a research direction that should be pursued to clarify
the nature of stochastic resetting and its consequences.

The ``local resetting'' process has recently been introduced in
an attempt to fill-in this gap \cite{Miron2021}. Conditioning the
success of a resetting attempt made by one degree of freedom, on the
states of all other degrees of freedom has already been applied \cite{Miron2021,Pelizzola_2021}
to study local resetting in a canonical model for interacting systems,
the simple exclusion model \cite{derrida2002large}. Since the renewal
approach is limited to global resetting, where all degrees of freedom
undergo resetting simultaneously, local resetting is inherently out
of its reach. Instead the mean field (MF) approximation was utilized
as a means of dealing with the correlations generated by introducing
interactions into the resetting process, and interesting observations
immediately followed: in the thermodynamic limit and for a fixed particle
density, it was shown that the stationary density profile becomes
entirely independent of the resetting rate, and that the profile's
shape scales with system size \cite{Miron2021}. It was also found
that introducing a system-size dependence into the resetting rate
has a non-trivial effect on the stationary profile's shape \cite{Miron2021,Pelizzola_2021}.
Another striking observation, which has yet to be explained, is the
existence of parameter regimes where density profiles computed using
the MF approximation yield an unexpectedly good fit to direct numerical
simulations \cite{Miron2021,Pelizzola_2021}. Local resetting evidently
opens an exciting door to the exploration of the interplay between
stochastic resetting and various types of interactions.

One ubiquitous setting which is known to generate non-trivial dynamics
in interacting systems is geometric confinement. The prototypical
setup for probing the consequence of geometric constraints on the
dynamics of interacting systems, is that of diffusive spherical particles
with repulsive interactions that are confined to a narrow channel.
Geometric confinement and repulsive interactions then typically generate
strong spatial and temporal correlations that make it difficult for
particles to overtake one-another. The resulting plethora of phenomena,
which have been investigated in numerous studies over the past several
decades, include tracer sub-diffusion, negative absolute and differential
mobility, and the vanishing of the velocity of a driven probe \cite{jepsen1965dynamics,percus1974anomalous,alexander1978diffusion,burlatsky1992directed,burlatsky1996motion,de1997dynamics,landim1998driven,benichou1999biased,candelier2010journey,illien2013active,cividini2016correlation,cividini2016exact,kundu2016exact,cividini2017driven,ahmadi2017diffusion,benichou2018unbinding,Miron_2020,Miron_2020_dyn,Miron2021_attraction}.
Following the recent introduction of local resetting, where interactions
enter the resetting process, one may wonder: ``how would local resetting
affect interacting systems whose dynamics are restricted by geometric
confinement?''.

Here we study the interplay between stochastic resetting and the geometrically
constrained dynamics of interacting diffusive particles. Envision
a narrow channel with rigid walls occupied by two species of diffusive
spherical particles that interact via short-ranged repulsion. Particles
of both species are completely identical except for two properties:
1) each species may have a different mass, and 2) the particles of
one of the species undergo local resetting, which here means that
an attempt to reset to a target location in the channel is successful
only if some fixed volume surrounding the target is vacant of any
other particle. Consider particles whose radius is not much smaller
than one quarter of the channel's width, where interactions and geometric
confinement conspire to reduce the rate at which neighboring particles
overtake one another. Analytically tackling this interacting non-equilibrium
system is a very challenging task with current methods and tools.
Yet its main ingredients, the interplay between local resetting and
geometric constraints, can be studied using a toy model that is more
susceptible to analytical treatment. Extending the efforts of \cite{Miron2021,Pelizzola_2021},
we model the narrow channel setup via the stochastic dynamics of hopping
particles on a $1D$ ring lattice of $L$ sites. We refer to particles
of the species undergoing local resetting as ``tracer'' particles,
while particles of the other species are called ``bath'' particles.
Hard-core exclusion interactions \cite{schutz1993phase,schutz1997exact,derrida1998exactly},
by which particles are excluded from entering an occupied site, replace
the channel's short-ranged repulsive interactions. The different diffusive
dynamics of each particle species in the channel, which follow from
their mass difference, are modeled by the different rates with which
particles attempt to hop into adjacent neighboring lattice sites,
setting the tracer hopping rate to $1$ and the bath particle hopping
rate to $D$. Tracers also undergo local resetting, independently
attempting to reset to the ``origin'' subject to exclusion. Finally,
the geometric confinement imposed by the narrow channel's walls is
modeled by an ``overtaking rate'' $\e$, at which a lattice particle
attempts to exchange places with a neighboring particle at an adjacent
site. 

After formulating equations for the evolution of the mean bath and
tracer particle density profiles, we apply the MF approximation and
compute the profiles in the stationary limit $t\ra\infty$. As in
the single species case \cite{Miron2021}, the stationary tracer density
profile $\t\left(x=\frac{\el}{L}\right)$ satisfies the same (stationary)
evolution equation as a single diffusive particle undergoing stochastic
resetting. $\t\left(x\right)$ correspondingly exhibits the typical
``tent-like'' shape \cite{Evans_2011}, with a cusp  at the origin
$x=0$. Yet the main result of this work concerns the behavior of
the stationary bath density profile $\r\left(x\right)$, whose behavior
changes dramatically as the ratio $\frac{\e}{D}$ crosses unity. For
$\frac{\e}{D}<1$ bath particles manage to escape the dense region
surrounding the origin, their density profile resembling an inverted
tent. But for $\frac{\e}{D}>1$ bath particles diffuse too slowly
to escape and a macroscopic number of them are ensnared near the origin,
resulting in a bath density profile whose shape is similar to $\t\left(x\right)$.
When the overtaking and bath hopping rates satisfy $\frac{\e}{D}=1$,
the bath profile becomes flat and equal to $\overline{\r}$. These
three regimes are predicted by MF theory and are validated by direct
numerical simulations of the lattice model's dynamics. While the MF
approximation becomes exact at $\e=D=1$ the agreement between the
MF and simulated profiles persists for $\e\approx D\approx1$. We
hope that the intriguing results that emerge through the interplay
of geometric confinement and local resetting will fuel experimental
and/or numerical investigations of the narrow channel setup and, more
generally, spark interest in how different types of interactions effect
the local resetting process.

The paper is organized as follows: Section \ref{sec:The-model} introduces
the $1D$ lattice model used to probe local resetting and geometric
confinement in the narrow-channel setup. Our main results, which consist
of analytical MF calculations and numerical simulations for the lattice
model, are presented in Sec. \ref{sec:Main-results}. Section \ref{sec:Evolution-equations-for}
presents the derivation of the evolution equations for the mean bath
and tracer density profiles, which contain correlations due to local
resetting and geometric confinement. In Sec. \ref{sec:Stationary-profiles-for}
we apply the MF approximation to the bath and tracer density profile
equations, solve them, and obtain the stationary profiles. The paper
is concluded in Sec. \ref{sec:Conclusion-and-discussion}, where we
discuss outstanding open questions and future prospects.

\section{The model \label{sec:The-model}}

\begin{figure}[H]
\begin{centering}
\includegraphics[scale=0.525]{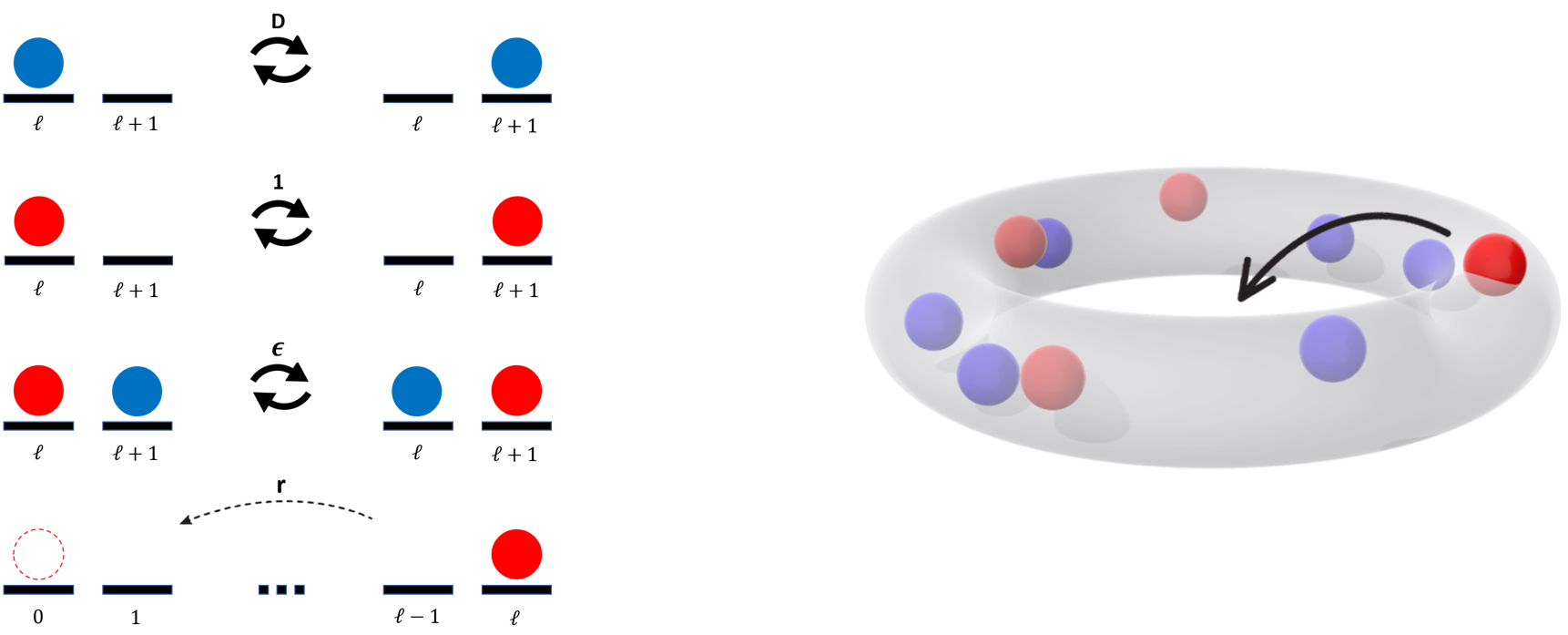}
\par\end{centering}
\caption{Illustration of the lattice dynamics and the narrow channel setup.
Bath particles are depicted in blue and resetting tracers in red.
\uline{Left}: Diagram detailing the $1D$ lattice dynamics. Bath
and tracer particles attempt hopping to adjacent sites with rates
$D$ and $1$ respectively, exchanges between neighboring particles
are attempted with rate $\protect\e$, and tracers attempt resetting
to the origin site $\protect\el=0$ with rate $r$. \uline{Right}:
Schematic illustration of the narrow channel setup with periodic boundaries,
showing a tracer in the midst of a resetting event. }

\label{illustration}
\end{figure}

We aim to construct a lattice model to describe the dynamics of two
species of diffusive spherical particles, locally resetting tracers
of mass $m_{t}$ and bath particles of mass $m_{b}$, that are confined
to a narrow periodic channel and subject to short-ranged repulsive
interactions (right panel of Fig. \ref{illustration}). To this end,
consider a $1D$ ring lattice of sites labeled $\el=0,...,L-1$ occupied
by $N$ bath particles and $M$ tracer particles, whose respective
mean densities $\overline{\r}=N/L$ and $\bar{\t}=M/L$ satisfy $\overline{\r}+\overline{\t}\le1$.
The system evolves in continuous time with the following stochastic
dynamics: a bath particle at site $\el$ attempts hopping to adjacent
neighboring sites $\el+1$ and $\el-1$ with rate $D$ to each side.
Tracer particles similarly attempt hopping to neighboring sites, but
with rate $1$ to each side. In addition to hopping, each tracer also
tries to reset its position to the ``origin'' site $\el=0$ with
rate $r$, independently of all other tracers, a process termed ``local
resetting'' in \cite{Miron2021}. Two mechanisms mediate interactions
between the particles. The first is hard-core ``exclusion'' \cite{schutz1993phase,schutz1997exact,derrida1998exactly},
by which hopping and resetting attempts that would lead a particle
into an occupied site are rejected. Correspondingly, each lattice
site can be occupied by one particle at most. The second mechanisms
is ``exchange'', where a particle at site $\el$ attempts with rate
$\e$ to exchange sites with a neighboring particle at sites $\el+1$
and $\el-1$. The attempt is rejected if the target site is vacant.
This mechanism has been previously used to model geometric constraints
in similar settings, including the narrow channel setup discussed
above \cite{cividini2017driven,cividini2018driven,Miron_2020,Miron_2020_dyn,Miron2021_attraction}.
Note that since particles of the same species are indistinguishable,
considering exchanges between two particles of the same species is
immaterial. Figure \ref{illustration} provides schematic illustrations
of the lattice dynamics (left panel) and the narrow channel setup
(right panel).

\section{Main results \label{sec:Main-results}}

\begin{figure}[H]
\begin{centering}
\includegraphics[scale=0.8]{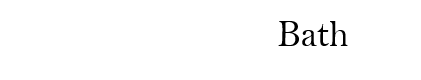} \hfill{} \includegraphics[scale=0.8]{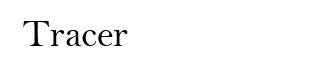}\vfill{}
\includegraphics[scale=0.55]{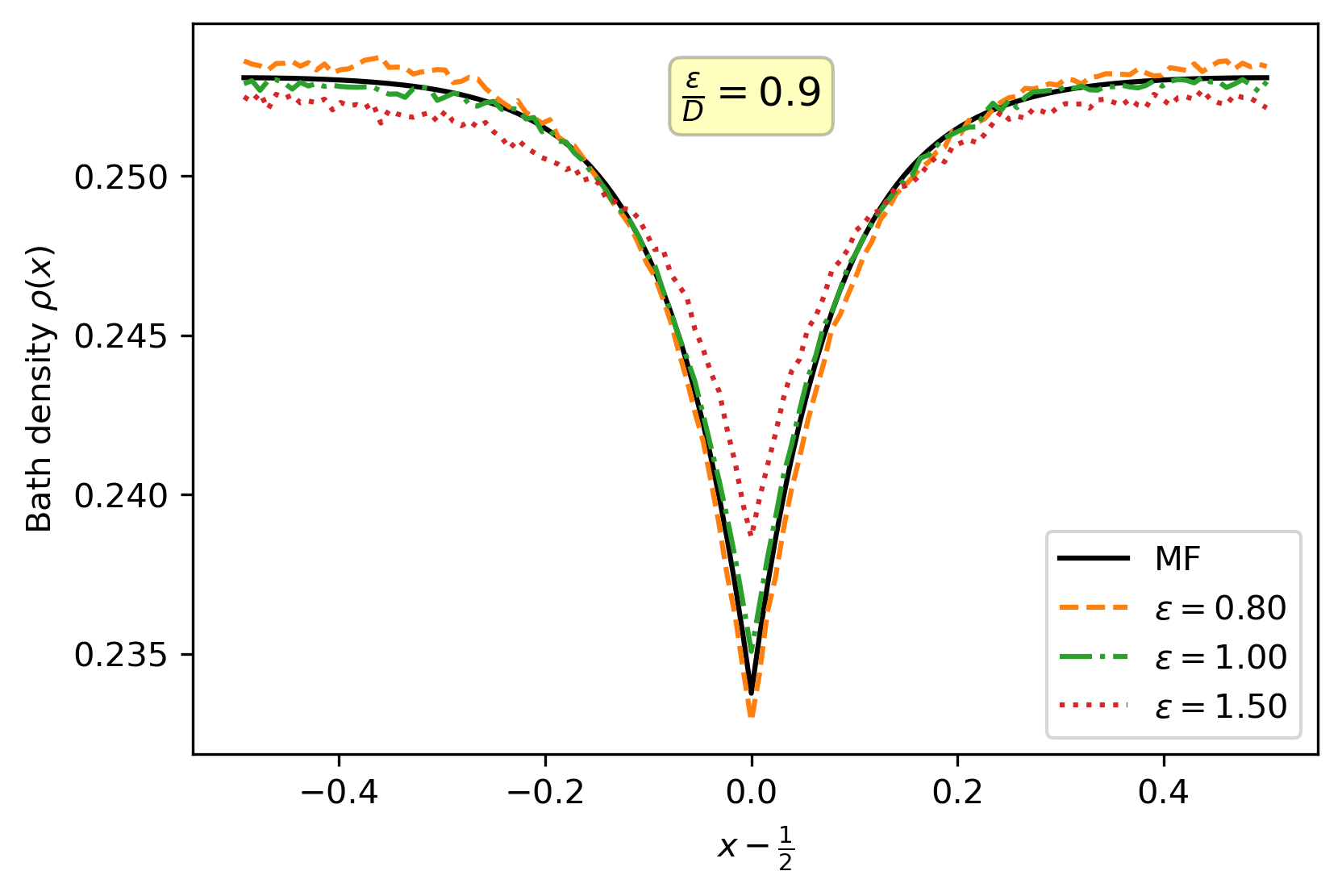} \hfill{} \includegraphics[scale=0.55]{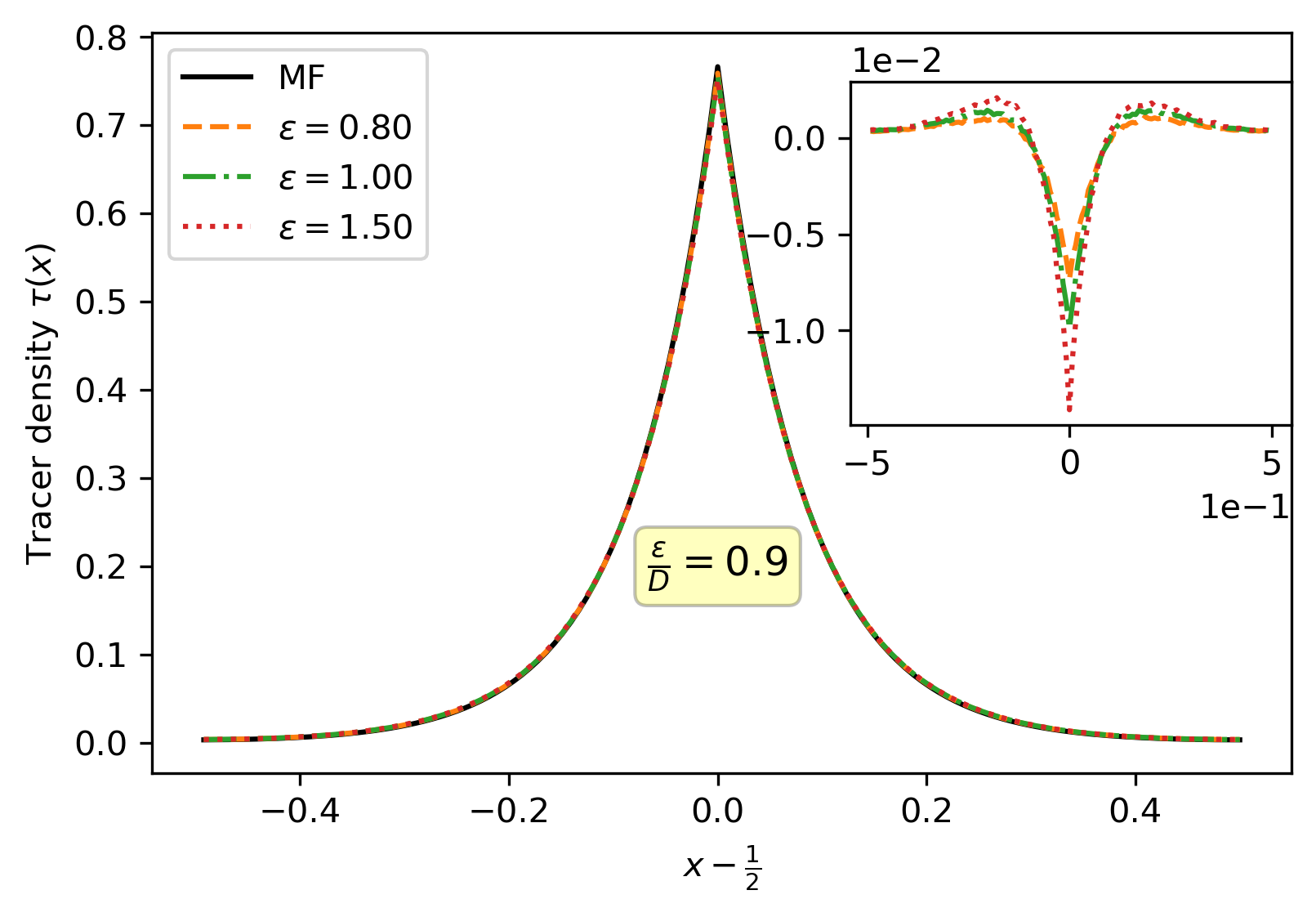}\vfill{}
\includegraphics[scale=0.55]{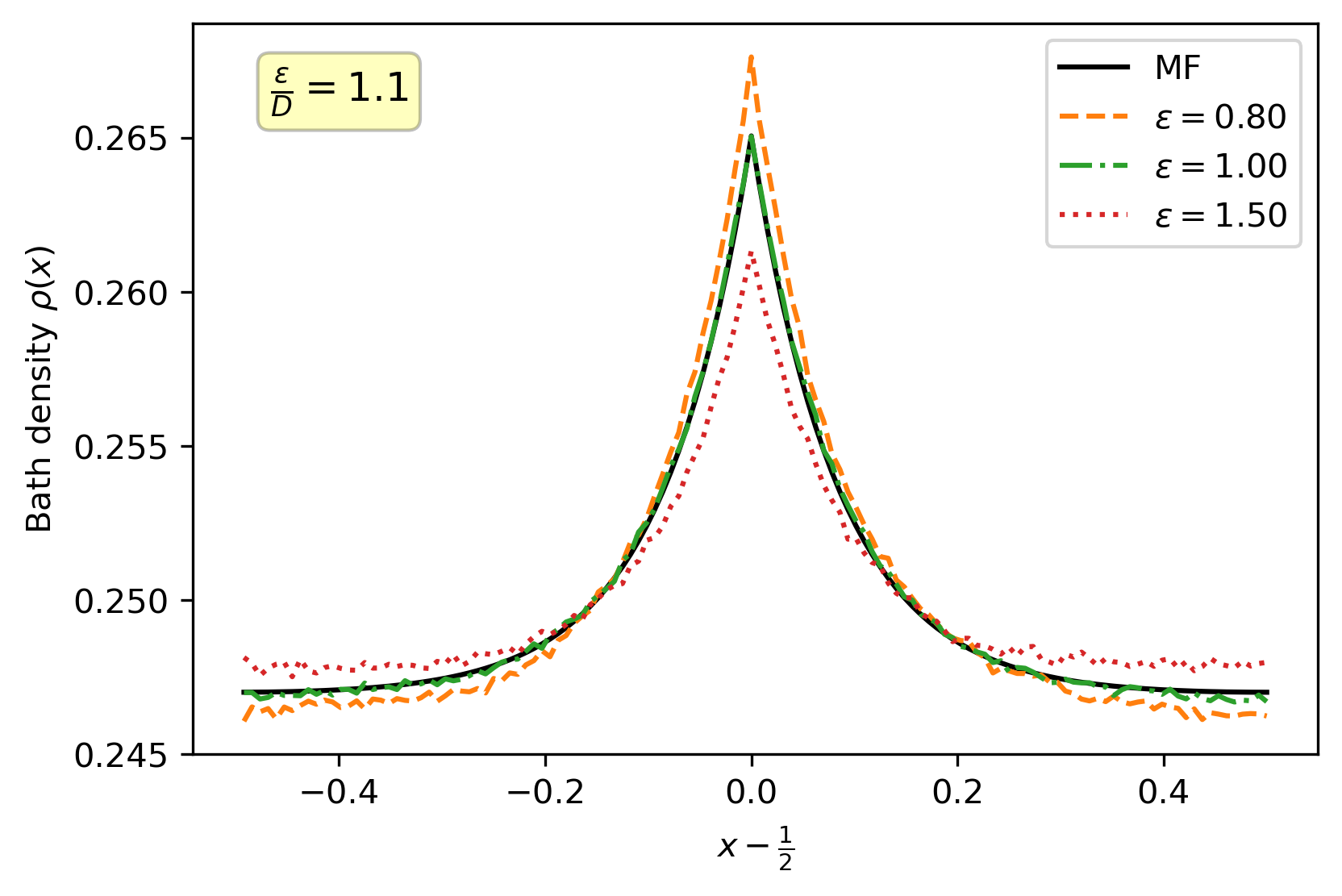} \hfill{} \includegraphics[scale=0.55]{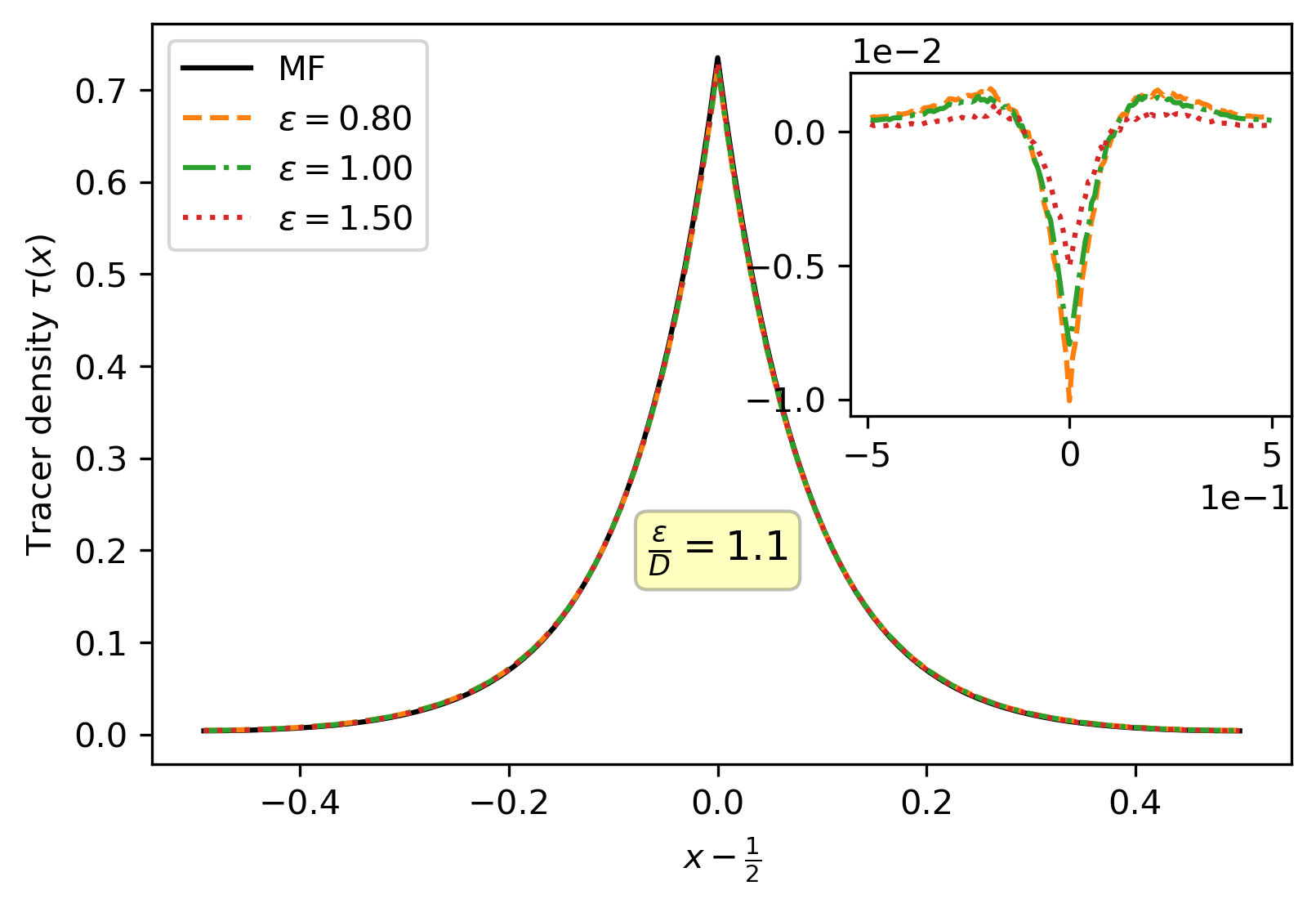}\vfill{}
\includegraphics[scale=0.55]{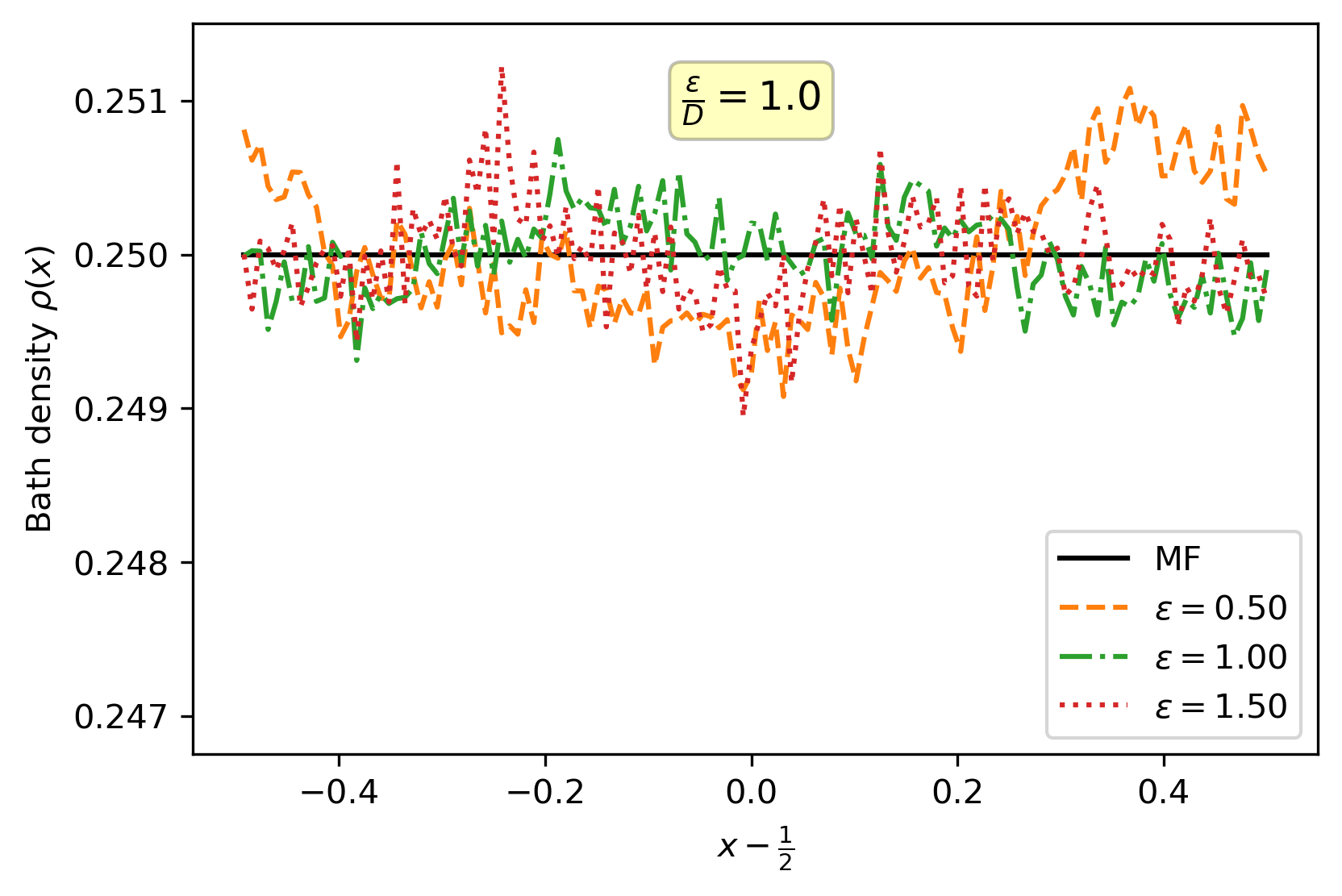} \hfill{} \includegraphics[scale=0.55]{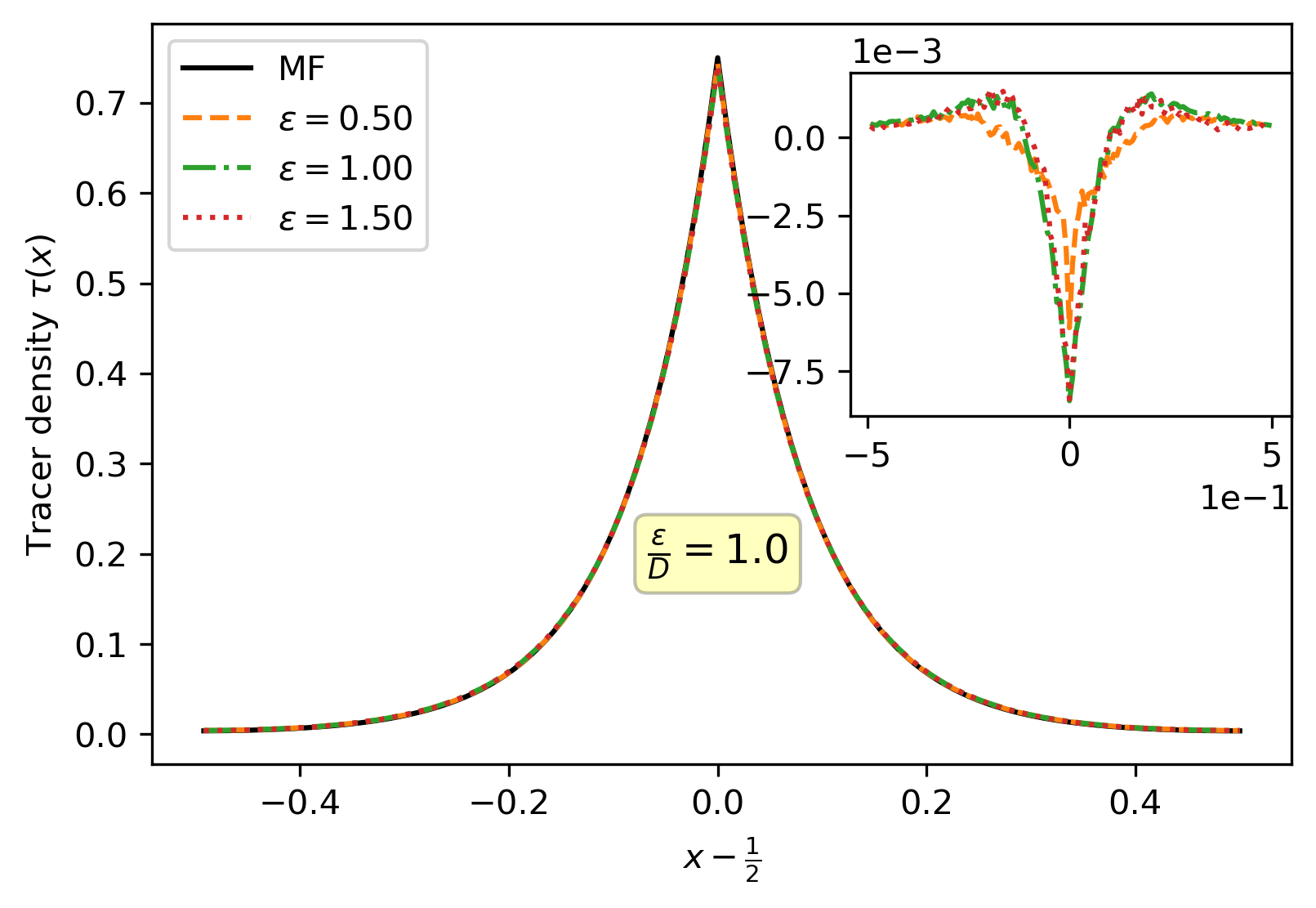}
\par\end{centering}
\caption{Stationary bath and tracer density profiles for $\overline{\protect\r}=1/4$,
$\overline{\protect\t}=1/8$, and $r=1$, with a fixed value of $\protect\e/D$
shown in each row. Notice that the horizontal axis was taken to be
$x-1/2$, which simply amounts to shifting the ``origin'' from $x=0$
to $x=1/2$, to highlight the behavior near the origin. Solid black
lines denote the MF profiles, computed in the thermodynamic limit
$L\protect\ra\infty$. In this limit Eqs. (\ref{eq:rho_tau_relation_fin}),
(\ref{eq:tau_profile}), and (\ref{eq:alpha eqn}), show that $\protect\e$
and $D$ only enter the MF profiles via their ratio $\protect\e/D$.
Dashed, dot-dashed, and dotted lines denote the simulated profiles,
obtained for $L=128$ and different values of $\protect\e$ and $D$.}
\label{profiles}
\end{figure}

The mean-field (MF) analysis and numerical simulation results presented
below demonstrate the non-trivial interplay between local resetting
and geometric confinement. In the thermodynamic limit the stationary
bath and tracer density profiles, $\r_{\el}$ and $\t_{\el}$ respectively,
are not functions of $\el$ and $L$ separately but are instead found
to depend on the scaling variable $x=\el/L\in\co{0,1}$, i.e. as $L\ra\infty$
we find $\r_{\el}\ra\r\left(x\right)$ and $\t_{\el}\ra\t\left(x\right)$. Moreover, in this limit, the total particle density at the origin approaches unity with finite-size corrections of $\sim\mathcal{O}\left(L^{-2}\right)$.
The tracer density profile $\t\left(x\right)$ is derived using the
MF approximation in Eq. (\ref{eq:tau_profile}) and features a tent-like
shape with a positive curvature $\t''\left(x\right)>0$ and a cusp
at the origin. This can be seen on the right column of Fig. \ref{profiles},
which compares the profiles obtained from MF and numerical simulations
for different values of $\e$ and $D$. We find that, through the
combined effect of local resetting and geometric confinement, a simple relation is generated between the bath and tracer profiles (see Eq.  (\ref{eq:rho_tau_relation_fin})), causing the
tent-like shape of $\t\left(x\right)$ to be imprinted on the stationary
density profile of the non-resetting bath particles $\r\left(x\right)$.
However, unlike the tracer profile $\t\left(x\right)$ which retains
its shape and curvature for any $\e,D\ne0$, the bath profile $\r\left(x\right)$
transitions \footnote{Note that this transition is not a ''phase transition'' in the classical sense, as no order-parameter becomes singular at this point.}  between ``repelled'' and ``trapped'' states as the
value of $\e/D$ crosses unity (see Eqs. (\ref{eq:rho_tau_relation_fin})
and (\ref{eq:tau_profile})). For $\e/D<1$ bath particles are strongly
repelled from the dense region near the origin, so that $\r\left(x\right)$
features a negative curvature $\r''\left(x\right)<0$ and the shape
of an inverted tent, as shown in the top left row of Fig. \ref{profiles}.
For $\e/D>1$ a finite fraction of bath particles remain trapped in
the dense region surrounding the origin. The left panel in the middle
row of Fig. \ref{profiles} shows that, in this case, $\r\left(x\right)$
has a positive curvature $\r''\left(x\right)>0$ and a shape similar
to that of $\t\left(x\right)$. At $\e/D=1$ the bath density profile
becomes completely flat, with its value equal to the mean bath density
$\overline{\r}$ (see left panel at the bottom row of Fig. \ref{profiles}).
The insets appearing throughout the right column of Fig. \ref{profiles}
plot the difference between the MF and simulated tracer profiles,
which is non-zero but still too small to see at scale.

The different states of $\r\left(x\right)$ are most easily understood
upon fixing the value of the exchange rate $\e$ to unity, thus setting
it equal to the tracer hopping rate. This limit corresponds to a channel
that is wide enough for the tracers' motion to be free of geometric
constraints over their diffusive time-scale. We start by considering
the repelled state $\e/D<1$, corresponding to $D>1$. In this state
tracers attempt hopping and exchanges with equal rates, but bath particles
hop at a higher rate. Imagine a bath particle trapped in the dense
region surrounding the origin. Clearly, it has a finite probability
to escape the dense region via exchanges with tracers. Once it succeeds
in doing so, reaching a region where there is a high density on one
side and a low density on the other, the bath particle is more likely
to distance itself by hopping away from the origin, rather than to
remain stuck therein. The opposite happens in the trapped state $\e/D>1$
(i.e. $D<1$), where the lower hopping rate implies a bath particle
has a higher chance to remain trapped in the dense region near the
origin. For $\e/D=1$ (i.e. $D=1$) the bath particles experience
a completely homogeneous environment, since hopping into a vacant
site and exchanging sites with a tracer occur with the same rate. 

\begin{figure}[H]
\begin{centering}
\includegraphics[scale=0.525]{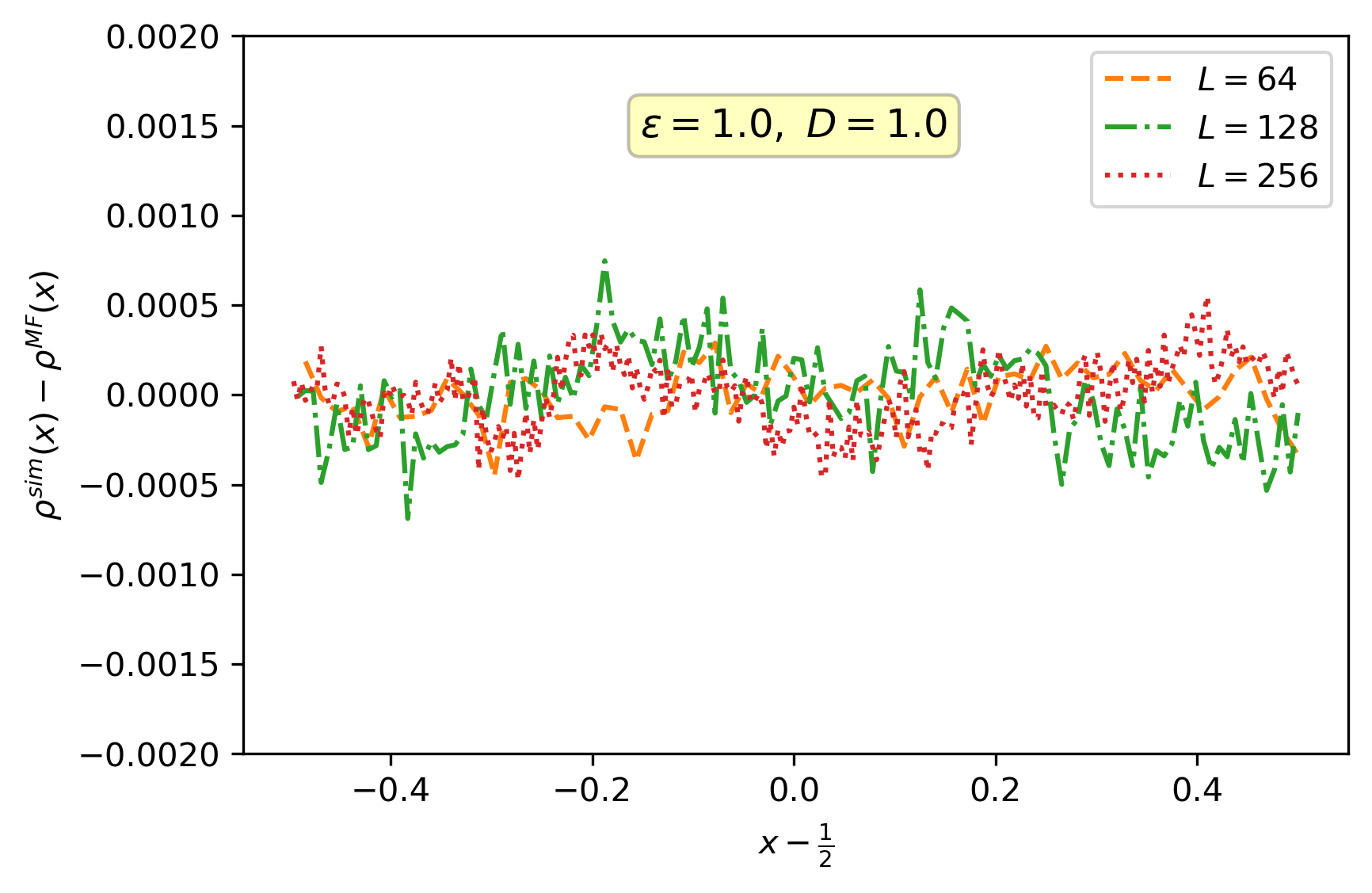} \hfill{} \includegraphics[scale=0.525]{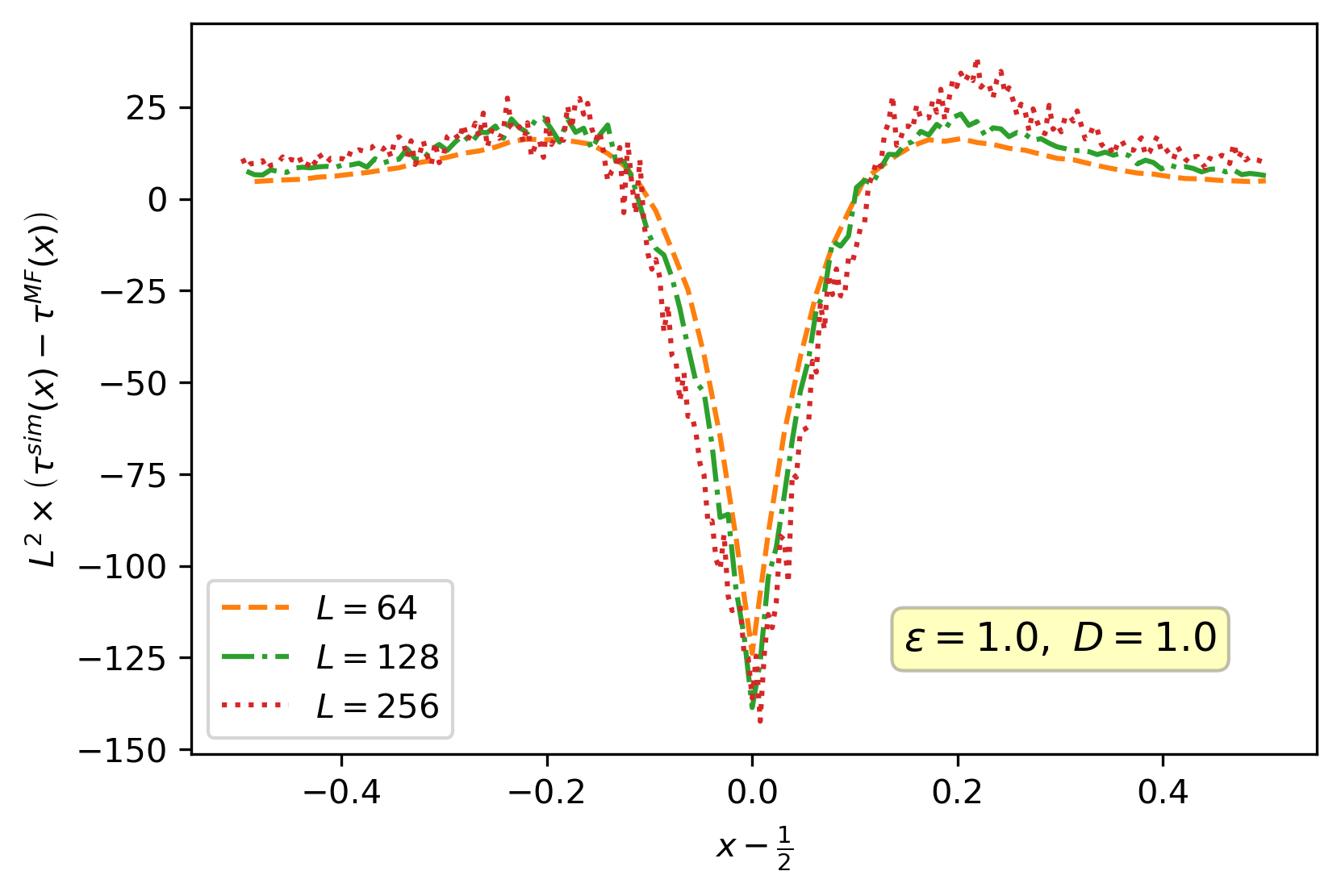}\vfill{}
\includegraphics[scale=0.525]{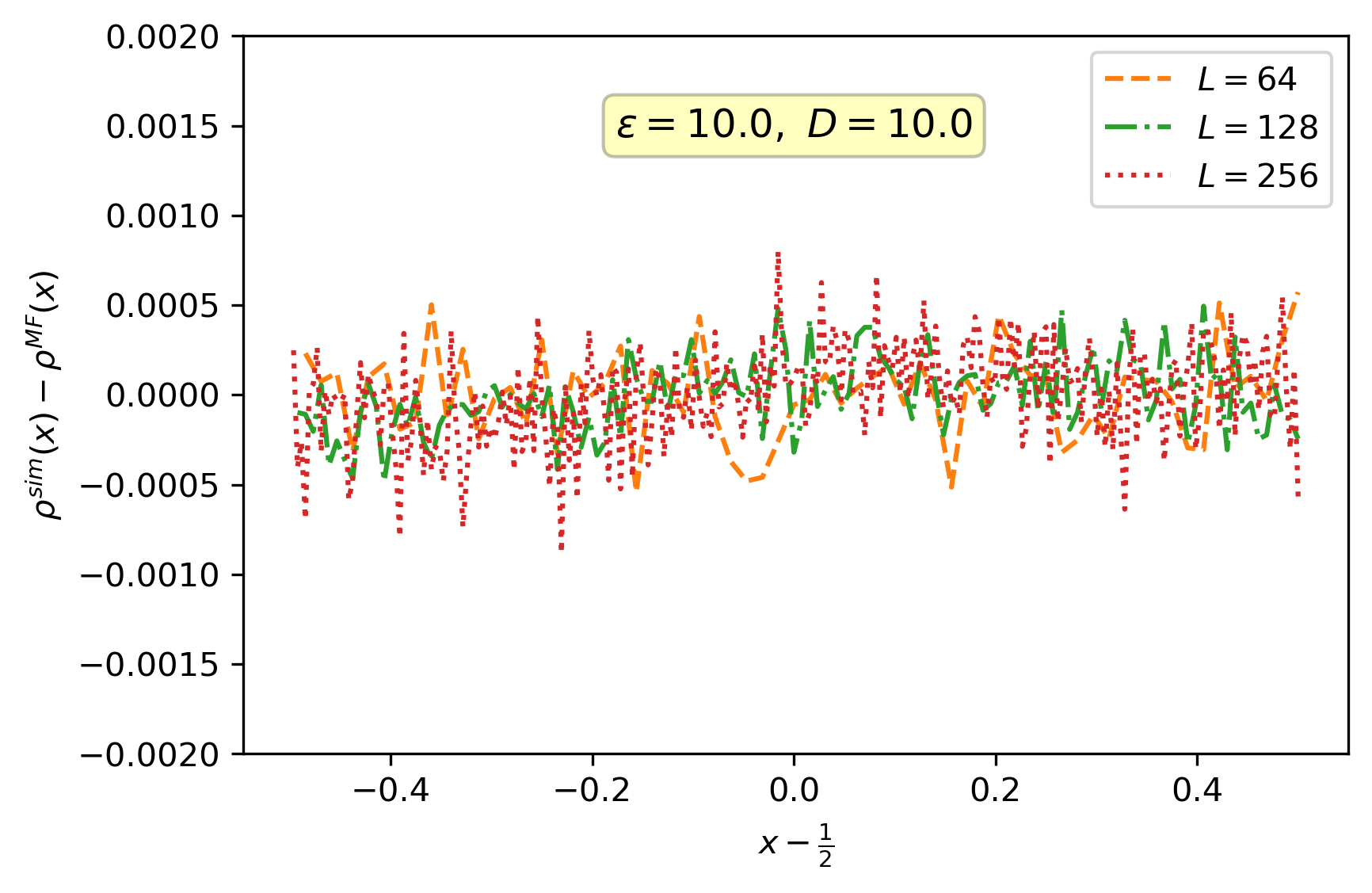} \hfill{} \includegraphics[scale=0.525]{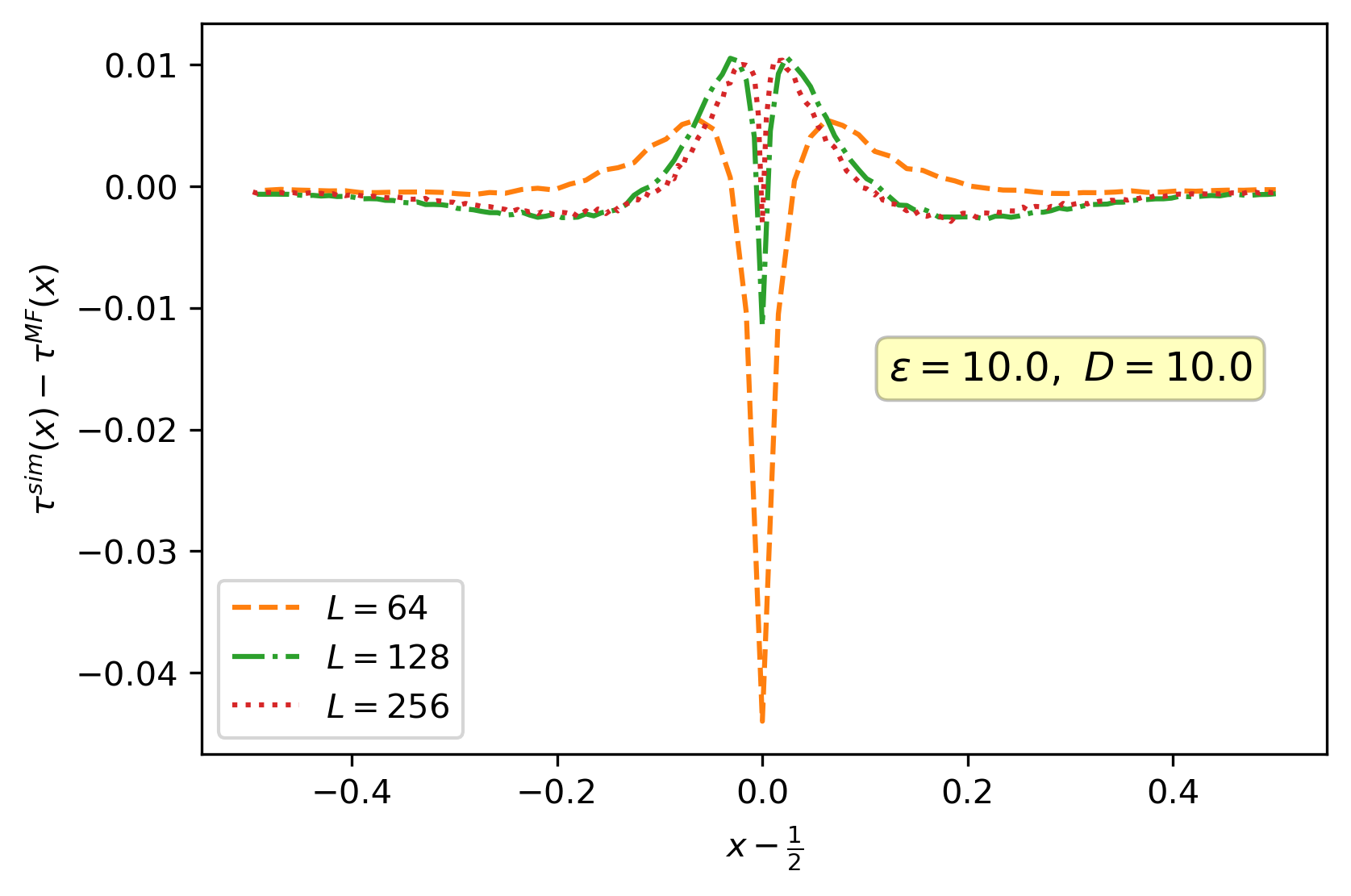}\vfill{}
\includegraphics[scale=0.525]{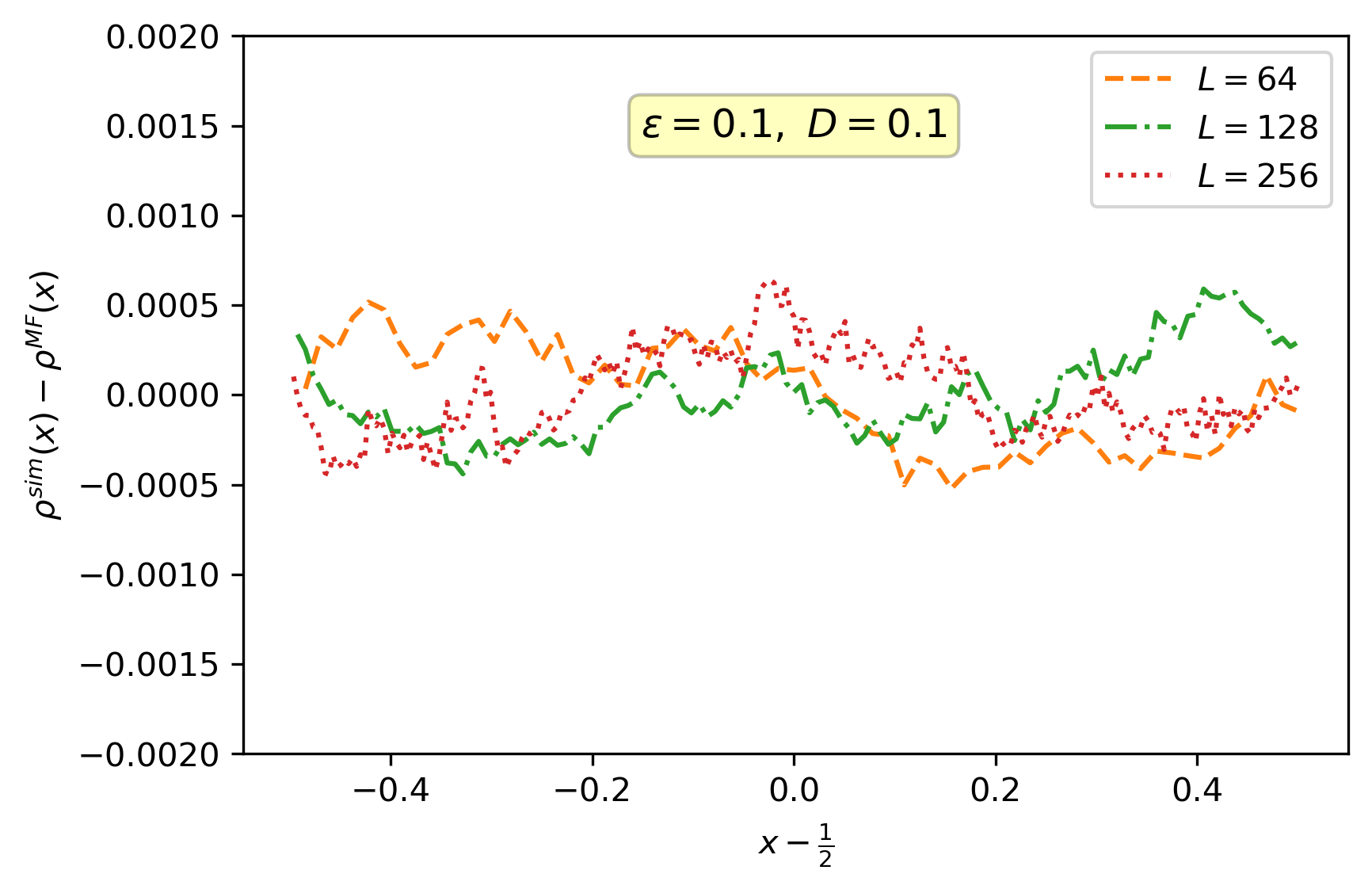} \hfill{} \includegraphics[scale=0.525]{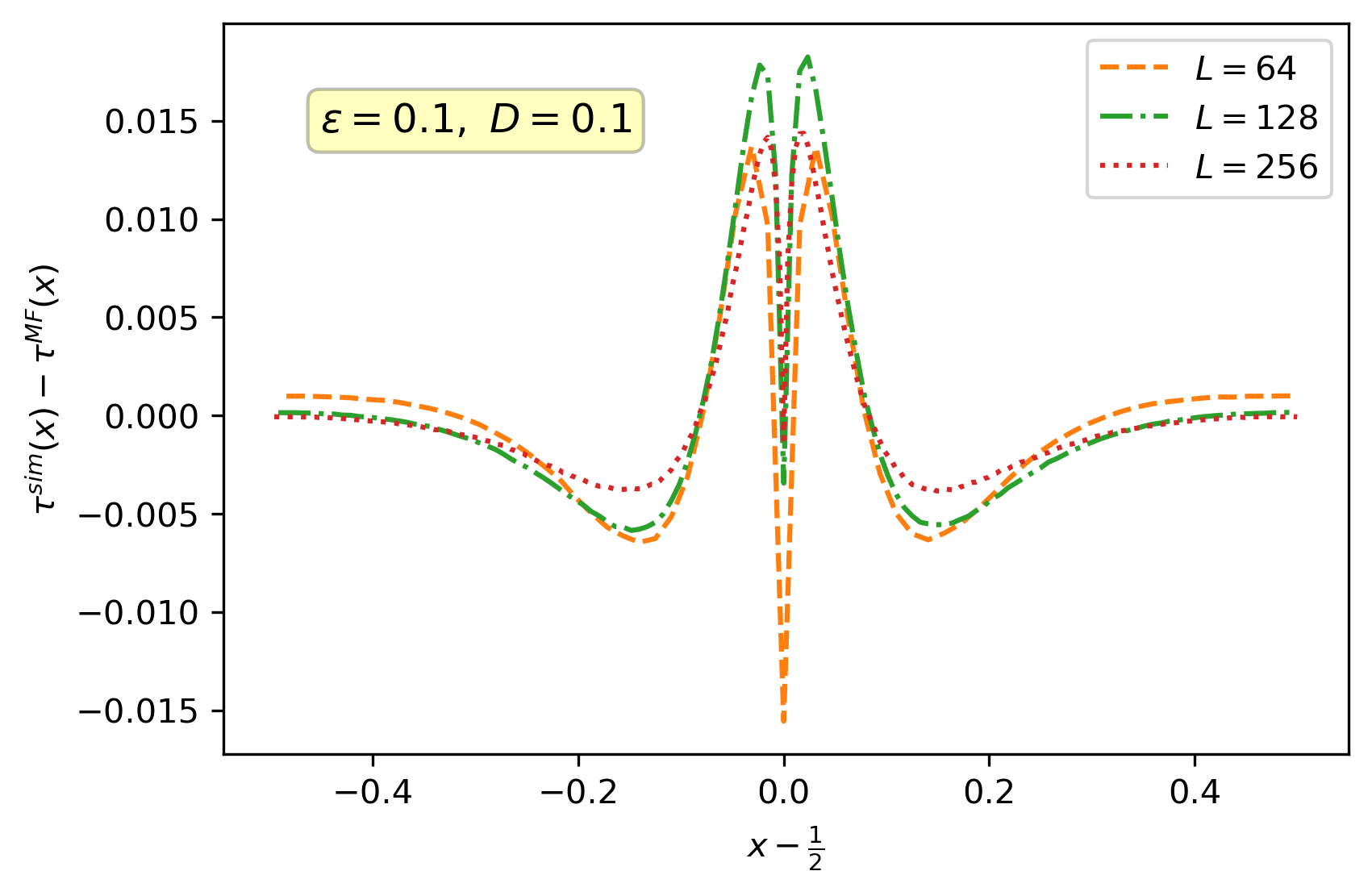}
\par\end{centering}
\caption{Data collapse of the difference between the simulated and MF profiles
for different system sizes $L=64,128,$ and $256$ for the parameters
$\overline{\protect\r}=1/4$, $\overline{\protect\t}=1/8$, $r=1$,
and $\protect\e/D=1$. Data for the bath and tracers particles is
presented on the left and right columns respectively. The first row
shows the data collapse for $\protect\e=D=1$, with the tracer density
difference scaled by $L^{2}$, i.e. $L^{2}\protect\d\protect\t\left(x\right)$.
The convincing collapse for $\protect\e=D=1$ implies that the MF
and simulated tracer density profiles become identical as $L\protect\ra\infty$.
For $\protect\e=D=10$ and $\protect\e=D=0.1$, shown in the second
and third rows, the tracer density difference is not scaled with system
size $L$ and the corrections to the MF profiles remain finite. Notice
that the horizontal axis is $x-1/2$, which simply amounts to shifting
the ``origin'' from $x=0$ to $x=1/2$. }
\label{differences eps/D =00003D 1}
\end{figure}

\begin{figure}[H]
\begin{centering}
\includegraphics[scale=0.525]{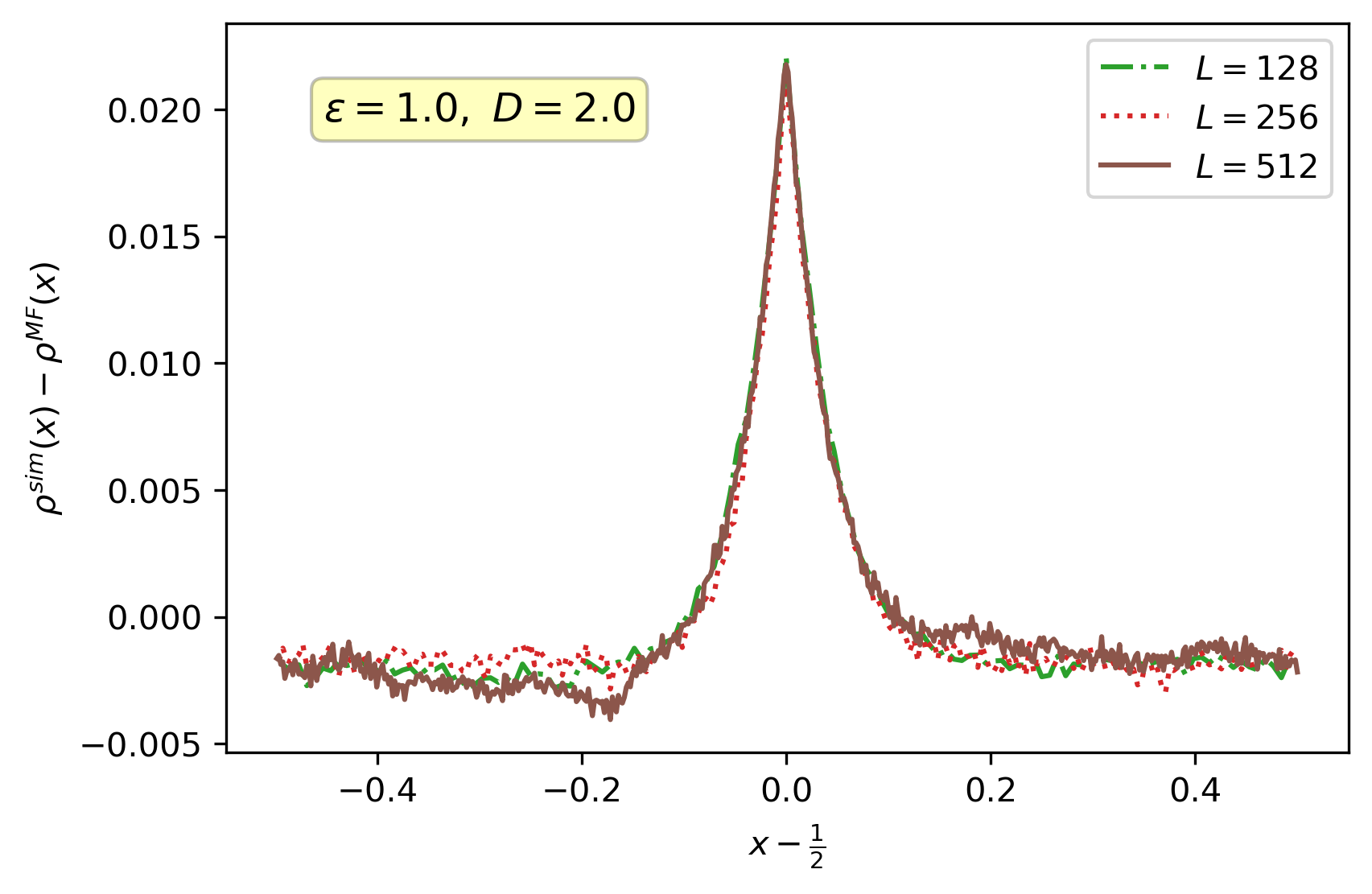} \hfill{} \includegraphics[scale=0.525]{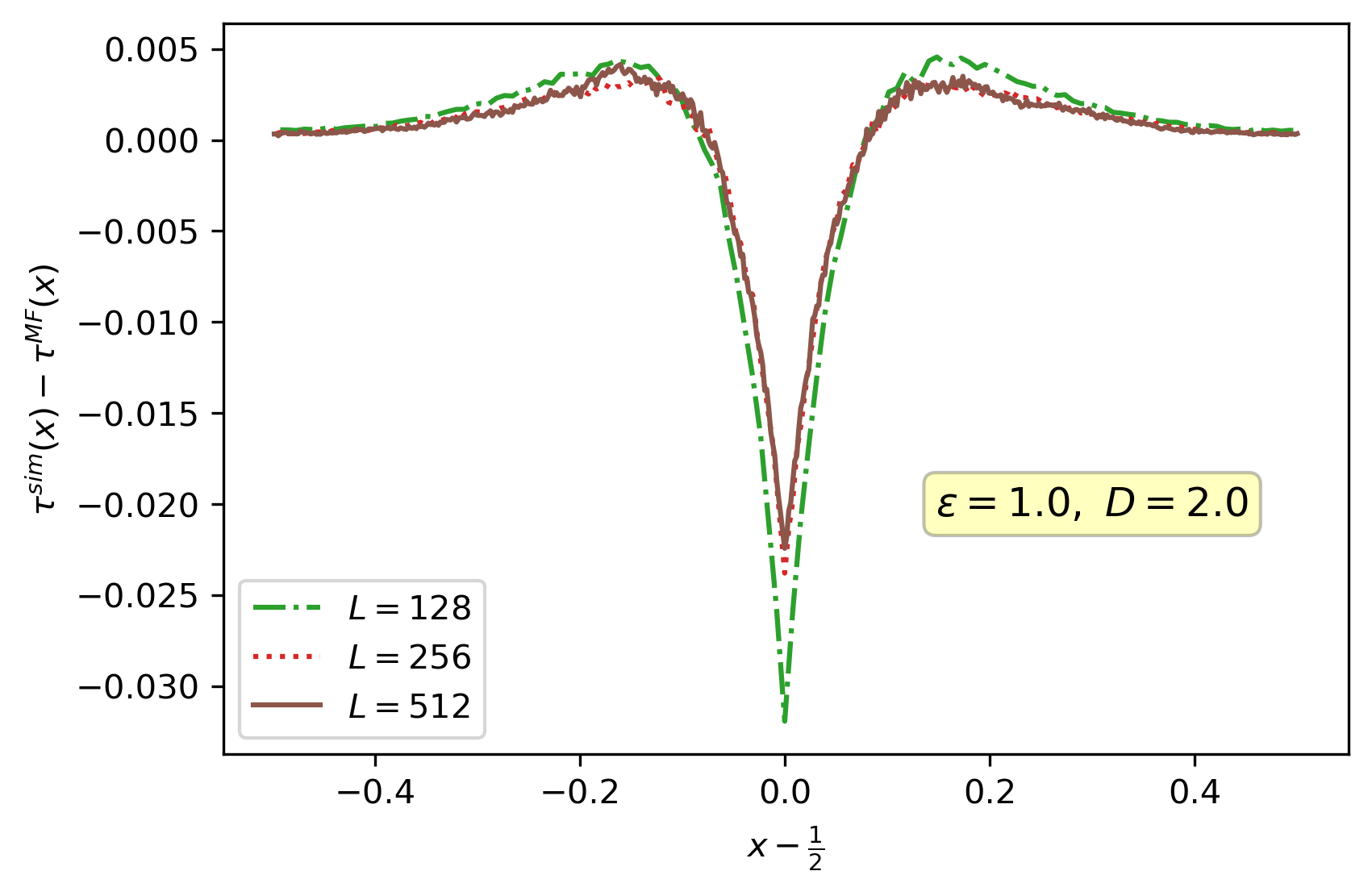}\vfill{}
\includegraphics[scale=0.525]{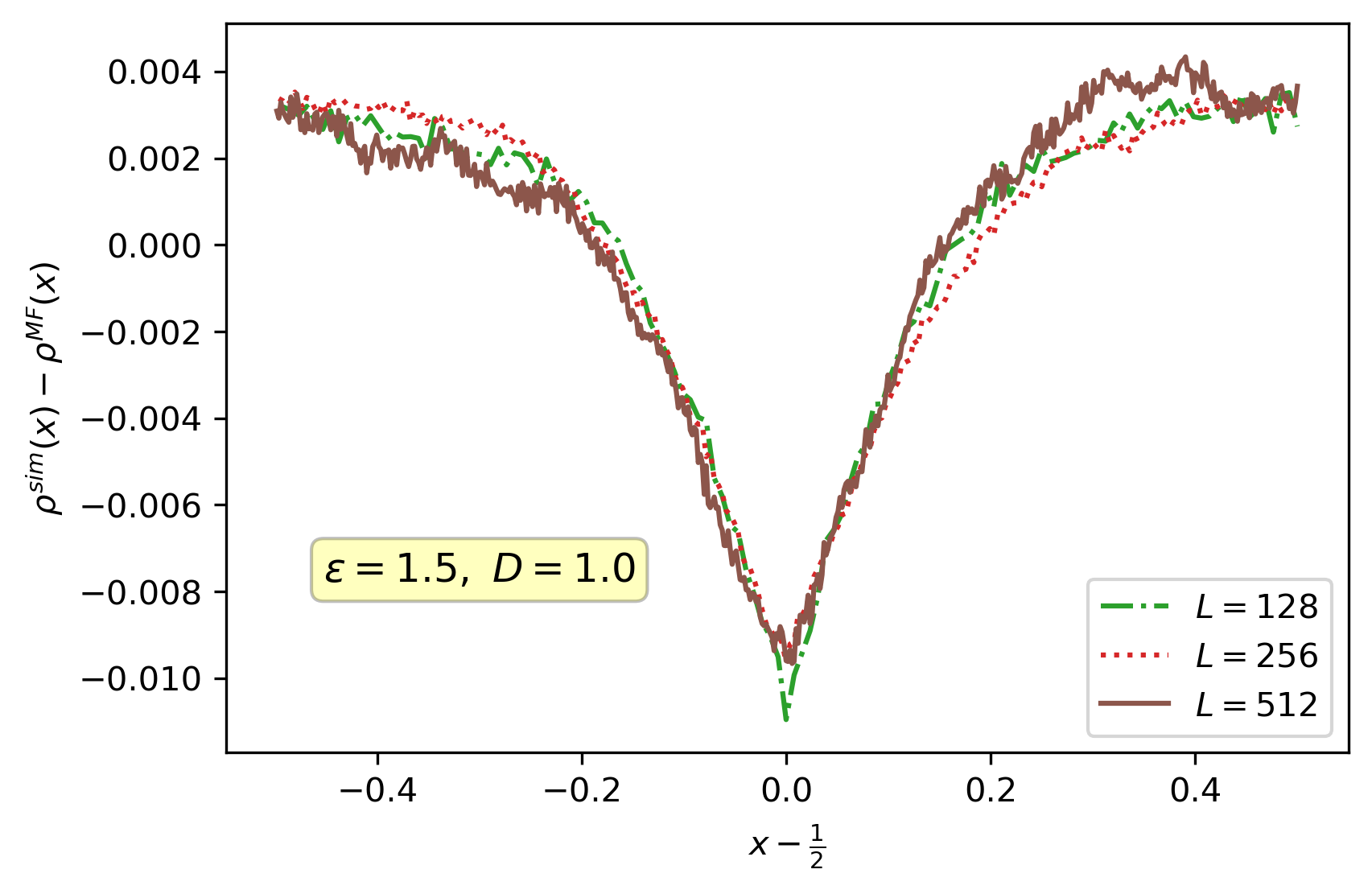} \hfill{} \includegraphics[scale=0.525]{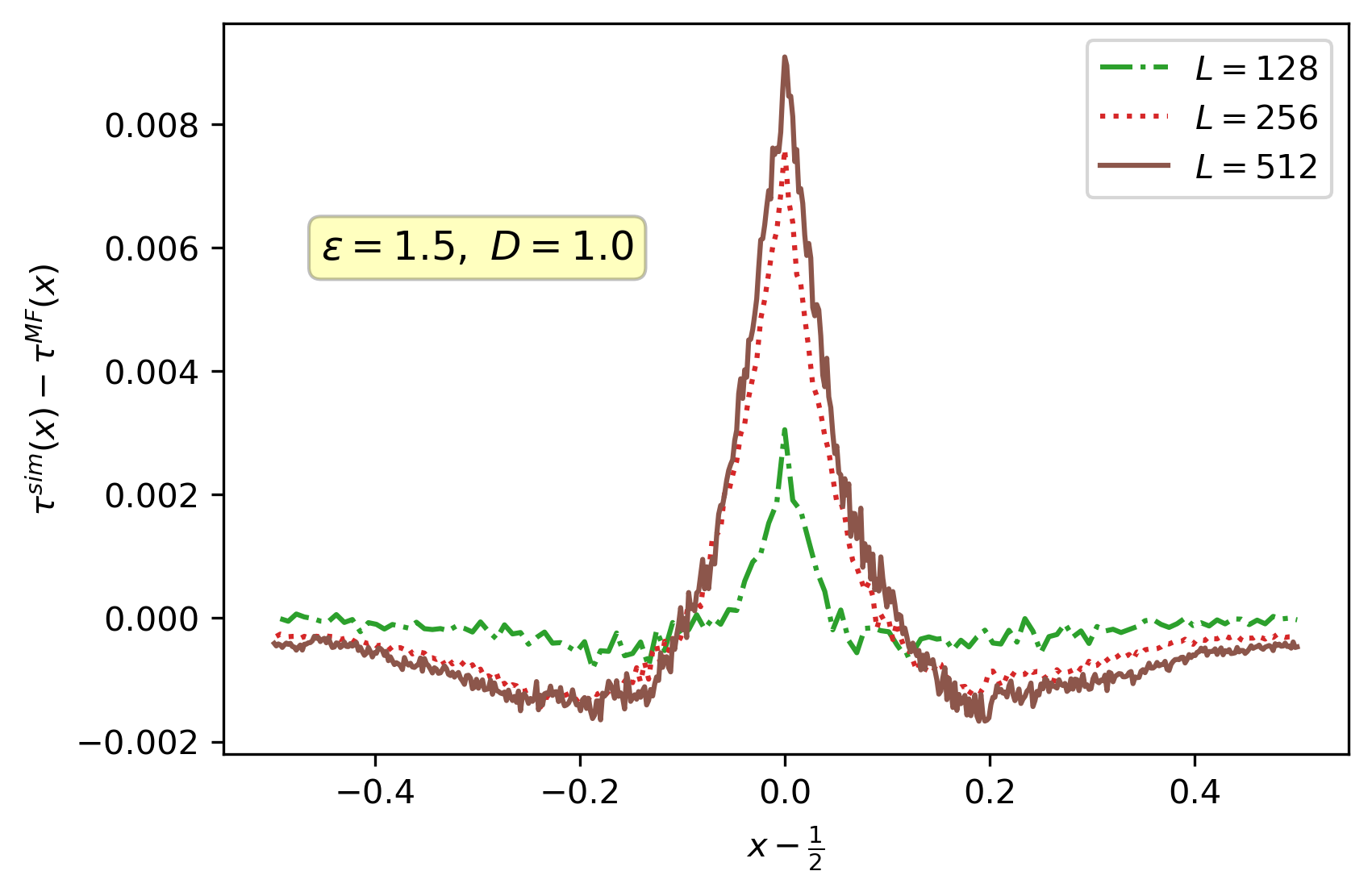}
\par\end{centering}
\caption{Data collapse of the difference between the simulated and MF profiles
for different system sizes $L=128,256,$ and $512$. The parameters
$\overline{\protect\r}=1/4$, $\overline{\protect\t}=1/8$, and $r=1$
are used throughout with $\protect\e/D\protect\ne1$. The first row
shows the collapse for $\protect\e=1$ and $D=2$, while the second
row is for $\protect\e=1.5$ and $D=1$. The collapse implies that
the MF and simulated density profiles remain different as $L\protect\ra\infty$.
Notice that the horizontal axis is $x-1/2$, which simply amounts
to shifting the ``origin'' from $x=0$ to $x=1/2$.}
\label{differences eps/D !=00003D 1}
\end{figure}

This leads us to the last main result of this study, which is that
the MF stationary density profiles becomes exact for $\e=D=1$. When
substituting $\e=D=1$ into Eqs. (\ref{eq:bath_den_simple}) for $\r\left(x\right)$
and $\t\left(x\right)$, the correlations due to the overtaking process
vanish from both equations. The equation for $\r\left(x\right)$ becomes
effectively decoupled from $\t\left(x\right)$, and is easily solved
to give $\overline{\r}$, but the remaining equation for $\t\left(x\right)$
still contains correlations due to the exchange process. For $\e=D=1$
the MF equation for $\t\left(x\right)$ is nearly identical to the
one studied in \cite{Miron2021} for locally resetting particles with
exclusion. There, for a fixed particle density and large $L$, the
remarkable agreement between the simulated and MF profiles was noted.
The same system was later studied numerically in \cite{Pelizzola_2021},
where the MF approximation was suggested to become exact in the limit
$L\ra\infty$. Following a numerical investigating the system we here
report that for $\e=D=1$ the MF approximation does indeed seem to
become exact in the thermodynamic limit $L\ra\infty$, despite the
existence of correlations coming from the local resetting process.
Figures \ref{differences eps/D =00003D 1} and \ref{differences eps/D !=00003D 1}
demonstrate this claim by showing data collapse of the difference
between the simulated and MF profiles $\d\r\left(x\right)\df\r^{sim}\left(x\right)-\r\left(x\right)$
and $\d\t\left(x\right)\df\t^{sim}\left(x\right)-\t\left(x\right)$
for different system sizes $L$, with each plot showing data for different
values of the parameters $\e$ and $D$. Data for $\e/D=1$ is shown
in Fig. \ref{differences eps/D =00003D 1} for $\e=D=0.1$, for $\e=D=10$,
and for the exact limit $\e=D=1$. For all parameter sets we find
$\d\r\left(x\right)\approx0$, up to structure-less noise due to finite
sampling. For $\e=D=1$ we observe that $L^{2}\times\d\t\left(x\right)$
exhibits a convincing data collapse, which suggests that $\d\t\left(x\right)\sim\ord{L^{-2}}$
and is consistent with the claim that corrections to MF decay as $L\ra\infty$.
Data collapse of $\d\t\left(x\right)$ for $\e=D=10$ and for $\e=D=0.1$
shows that the correction to MF remain finite as $L\ra\infty$. We
complement this picture in Fig. \ref{differences eps/D !=00003D 1},
which shows data collapse for $\e=1$ and $D=2$ (i.e. $\e/D<1$),
and for $\e=1.5$ and $D=1$ (i.e. $\e/D>1$). Here the resulting
differences $\d\r\left(x\right)$ and $\d\t\left(x\right)$ are small
but also remain finite as $L$ increases. Notice that, as in Fig.
\ref{profiles}, the horizontal axis in Figs. \ref{differences eps/D =00003D 1}
and \ref{differences eps/D !=00003D 1} was set to $x-1/2$ to ensure
the behavior near the origin is easily and clearly visible. 

A final comment about the model parameters is in order. In an effort
to isolate the consequence of changing only a subset of the model's
parameters at a time, the figures described above all correspond to
the mean bath density $\overline{\r}=1/8$, mean tracer density $\overline{\t}=1/4$,
and resetting rate $r=1$. Nevertheless, the validity of the entailing
results has been confirmed for a variety of mean densities and values
of $r$. 

\section{Evolution equations for the mean density profiles \label{sec:Evolution-equations-for}}

We proceed by formulating the equations governing the evolution of
the mean bath and tracer density profiles. Let $\tilde{\r}_{\el}\left(t\right)$
denote the instantaneous occupation of site $\el$ by a bath particle
at time $t$, assuming the value $1$ if a bath particle is present
and $0$ otherwise, and let $\tilde{\t}_{\el}\left(t\right)$ denote
the same for tracers. Considering the configurations affecting the
occupation of a bath particle at site $\el$ and averaging over the
system's stochastic dynamics, we obtain an evolution equation for
the mean bath density profile $\left\langle \tilde{\r}_{\el}\left(t\right)\right\rangle $

\[
\partial_{t}\left\langle \tilde{\r}_{\el}\right\rangle =D\left\langle \left(1-\tilde{\r}_{\el}-\tilde{\t}_{\el}\right)\left(\tilde{\r}_{\el+1}+\tilde{\r}_{\el-1}\right)\right\rangle -D\left\langle \tilde{\r}_{\el}\left[\left(1-\tilde{\r}_{\el+1}-\tilde{\t}_{\el+1}\right)+\left(1-\tilde{\r}_{\el-1}-\tilde{\t}_{\el-1}\right)\right]\right\rangle 
\]
\begin{equation}
+\e\left\langle \tilde{\t}_{\el}\left(\tilde{\r}_{\el+1}+\tilde{\r}_{\el-1}\right)\right\rangle -\e\left\langle \tilde{\r}_{\el}\left(\tilde{\t}_{\el+1}+\tilde{\t}_{\el-1}\right)\right\rangle .\label{eq:bath_den}
\end{equation}
The first term on the right-hand side describes the bath density gain
at site $\el$ due to configurations where a bath particle at site
$\el\pm1$ attempts hopping to site $\el$ with rate $D$. Notice
the ``exclusion factor'' $\left(1-\tilde{\r}_{\el}\left(t\right)-\tilde{\t}_{\el}\left(t\right)\right)$
weighing this term, which vanishes if site $\el$ is already occupied
by a bath or tracer particle, and is $1$ otherwise. The second term
describes the loss of bath density due to the bath particle attempting
to leave site $\el$ by hopping into an adjacent site. The third and
fourth terms analogously describe gain and loss terms associated with
exchanges between bath and tracer particles, which are attempted with
rate $\e$. The equation for the mean tracer density is 
\[
\partial_{t}\left\langle \tilde{\t}_{\el}\right\rangle =\left\langle \left(\tilde{\t}_{\el+1}+\tilde{\t}_{\el-1}\right)\left(1-\tilde{\r}_{\el}-\tilde{\t}_{\el}\right)\right\rangle -\left\langle \tilde{\t}_{\el}\left[\left(1-\tilde{\r}_{\el+1}-\tilde{\t}_{\el+1}\right)+\left(1-\tilde{\r}_{\el-1}-\tilde{\t}_{\el-1}\right)\right]\right\rangle 
\]
\begin{equation}
+\e\left\langle \tilde{\r}_{\el}\left(\tilde{\t}_{\el+1}+\tilde{\t}_{\el-1}\right)\right\rangle -\e\left\langle \tilde{\t}_{\el}\left(\tilde{\r}_{\el+1}+\tilde{\r}_{\el-1}\right)\right\rangle +\tilde{\c}_{\el},\label{eq:tracer_den}
\end{equation}
and its construction is analogous to that of Eq. (\ref{eq:bath_den}).
The two equations have nearly the same structure, with the exception
of the last term $\tilde{\c}_{\el}$ on the right-hand side of Eq.
(\ref{eq:tracer_den}). This term, which describes the local resetting
of tracers
\begin{equation}
\tilde{\c}_{\el}=\begin{cases}
\sum_{n=1}^{L-1}r\left\langle \left(1-\tilde{\r}_{0}-\tilde{\t}_{0}\right)\tilde{\t}_{n}\right\rangle  & \el=0\\
-r\left\langle \left(1-\tilde{\r}_{0}-\tilde{\t}_{0}\right)\tilde{\t}_{\el}\right\rangle  & \el\ne0
\end{cases},\label{eq:chi_tilde}
\end{equation}
behaves differently at site $\el=0$ versus sites $\el\ne0$: $\tilde{\c}_{\el=0}$
is a gain term at site $\el=0$ that accounts for the attempts made
by tracers, located anywhere besides the origin, to reset their position
to the origin with rate $r$, whereas $\tilde{\c}_{\el\ne0}$ describes
the loss of tracer density at sites $\el\ne0$ due to resetting attempts
with rate $r$. The exclusion factor $\left(1-\tilde{\r}_{0}-\tilde{\t}_{0}\right)$
in $\tilde{\c}_{\el}$ ensures that resetting attempts are rejected
if the origin is occupied.

Defining the lattice Laplacian $\grad_{\el}^{2}$ and gradient $\grad_{\el}$
operators, which act on a lattice variable $u_{\el}$ as $\grad_{\el}^{2}u_{\el}=u_{\el+1}-2u_{\el}+u_{\el-1}$
and $\grad_{\el}u_{\el}=u_{\el}-u_{\el-1}$, and respectively denoting
the average bath and tracer density profiles by $\r_{\el}=\left\langle \tilde{\r}_{\el}\right\rangle $
and $\t_{\el}=\left\langle \tilde{\t}_{\el}\right\rangle $ simplifies
Eqs. (\ref{eq:bath_den}) and (\ref{eq:tracer_den}) to
\begin{equation}
\frac{1}{D}\partial_{t}\r_{\el}=\grad_{\el}^{2}\r_{\el}+\left(1-\frac{\e}{D}\right)\grad_{\el}\left(\left\langle \tilde{\r}_{\el}\tilde{\t}_{\el+1}\right\rangle -\left\langle \tilde{\t}_{\el}\tilde{\r}_{\el+1}\right\rangle \right),\label{eq:bath_den_simple}
\end{equation}
and 
\begin{equation}
\partial_{t}\t_{\el}=\grad_{\el}^{2}\t_{\el}-\left(1-\e\right)\grad_{\el}\left(\left\langle \tilde{\r}_{\el}\tilde{\t}_{\el+1}\right\rangle -\left\langle \tilde{\t}_{\el}\tilde{\r}_{\el+1}\right\rangle \right)+\tilde{\c}_{\el}.\label{eq:tracer_den_simple}
\end{equation}
It is now apparent that the mean density profiles $\r_{\el}$ and
$\t_{\el}$ are affected by correlations that separate into two contributions:
``exchange correlations'', which appear in the second term on the
right-hand sides of both equations and are generated by the exchange
interactions, and ``local resetting'' correlations, which only enter
Eq. (\ref{eq:tracer_den_simple}) via $\tilde{\c}_{\el}$ and are
generated by the local resetting process. Particle conservation places
an additional constraint on $\r_{\el}$ and $\t_{\el}$, requiring
that 
\begin{equation}
\sum_{\el=0}^{L-1}\r_{\el}\left(t\right)=N\text{ and }\sum_{\el=0}^{L-1}\t_{\el}\left(t\right)=M.\label{eq:conservation_0}
\end{equation}

Notice that the exchange correlations vanish from both equations for
$\e=D=1$. In this case at sites $\el\ne0$ (in the ``bulk'' of
the system) Eqs. (\ref{eq:bath_den_simple}) and (\ref{eq:tracer_den_simple})
reduce to
\begin{equation}
\partial_{t}\r_{\el}=\grad_{\el}^{2}\r_{\el},\label{eq:trivial rho eqn-1}
\end{equation}
and 
\begin{equation}
\partial_{t}\t_{\el}=\grad_{\el}^{2}\t_{\el}-r\left\langle \left(1-\tilde{\r}_{0}-\tilde{\t}_{0}\right)\tilde{\t}_{\el}\right\rangle ,\label{eq:trivial tau eqn-1}
\end{equation}
where the last term in the $\t_{\el}$ equation comes from Eq. (\ref{eq:chi_tilde})
for $\el\ne0$. With the effect of correlations eliminated from Eq.
(\ref{eq:trivial rho eqn-1}) for $\r_{\el}$, we can now solve it
independently of Eq. (\ref{eq:trivial tau eqn-1}) for $\t_{\el}$.
Imposing periodic boundary conditions and bath particle number conservation
$\sum_{\el=0}^{L-1}\r_{\el}=N$ yields the stationary solution $\overline{\r}$.
Equation (\ref{eq:trivial tau eqn-1}) for $\t_{\el}$ is nearly identical
to the one considered in \cite{Miron2021}, which studied identical
particles undergoing diffusion and local resetting. For a finite particle
density and large $L$, the stationary density profile obtained there
using the MF approximation was found to be remarkably similar to the
empirical profile obtained via numerical simulations of the model's
dynamics, a result which was later validated in \cite{Pelizzola_2021}.
Together with the support provided by the numerical evidence presented
in Sec. \ref{sec:Main-results}, when local resetting is the only
source for correlations, as in Eqs. (\ref{eq:trivial tau eqn-1})
and (\ref{eq:trivial rho eqn-1}), the MF approximation seems to become
exact in the thermodynamic limit. Applying the MF approximation to
the ``full'' Eqs. (\ref{eq:trivial rho eqn-1}) and (\ref{eq:trivial tau eqn-1}),
which contain general overtaking rates and general bath and tracer
diffusion rates, is correspondingly expected to provide an increasingly
accurate description of the model as $\e,D\ra1$. We comment that for $\e=1$ but a general $D \ne 1$, Eq. (\ref{eq:tracer_den_simple}) still contains correlations between the tracer occupation at site $\el$ and the bath occupation at the origin.

\section{Stationary mean-field profiles for general $\protect\e$ and $D$
\label{sec:Stationary-profiles-for}}

Equations (\ref{eq:bath_den_simple}) and (\ref{eq:tracer_den_simple})
for the average density profiles $\r_{\el}$ and $\t_{\el}$ involve
correlations between the instantaneous particle occupations. These
pair of equations are actually the first members of an extensive hierarchy
of partial difference equations for all moments of the full $N$ bath
particle and $M$ tracer particle distribution, which one would have
to formulate and solve\footnote{One may apply the procedure used in deriving Eqs. (\ref{eq:bath_den})
and (\ref{eq:tracer_den}) for $\left\langle \tilde{\r}_{\el}\right\rangle $
and $\left\langle \tilde{\t}_{\el}\right\rangle $ to formulate equations
for the two-point correlations sitting at the next level of the hierarchy,
as well as any higher order correlations.} to rigorously account for the effect of interactions between the
particles on the density profiles. Unfortunately, this hierarchy quickly
becomes intractable and there are only a handful of models for which
exact methods are known to apply \cite{blythe2007nonequilibrium,kriecherbauer2010pedestrian}.
To make progress we thus resort to the MF approximation, which essentially
sets all cumulants to zero, so that $n$-point correlations factorize
into products (e.g. $2$-point functions become $\left\langle \tilde{u}_{n}\tilde{v}_{m}\right\rangle \ra\left\langle \tilde{u}_{n}\right\rangle \left\langle \tilde{v}_{\el}\right\rangle $).
Applying the MF approximation to Eqs. (\ref{eq:bath_den_simple})
and (\ref{eq:tracer_den_simple}) yields 
\begin{equation}
\frac{1}{D}\partial_{t}\r_{\el}=\grad_{\el}^{2}\r_{\el}+\left(1-\frac{\e}{D}\right)\grad_{\el}\left(\r_{\el}\t_{\el+1}-\t_{\el}\r_{\el+1}\right),\label{eq:bath_den_simple-1}
\end{equation}
and 
\begin{equation}
\partial_{t}\t_{\el}=\grad_{\el}^{2}\t_{\el}-\left(1-\e\right)\grad_{\el}\left(\r_{\el}\t_{\el+1}-\t_{\el}\r_{\el+1}\right)+\c_{\el},\label{eq:tracer_den_simple-1}
\end{equation}
where $\tilde{\c}_{\el}$ in Eq. (\ref{eq:tracer_den_simple}) is
replaced by 
\begin{equation}
\c_{\el}=\begin{cases}
\sum_{n=1}^{L-1}r\left(1-\r_{0}-\t_{0}\right)\t_{n} & \el=0\\
-r\left(1-\r_{0}-\t_{0}\right)\t_{\el} & \el\ne0
\end{cases}.\label{eq:chi}
\end{equation}
Despite the apparent crudeness of uncontrollably factorizing correlations,
as shown in Sec. \ref{sec:Main-results}, comparing the MF solutions
to numerical simulation results suggest that MF effectively captures
the salient features arising from the interplay between geometrically
constrained diffusion and local resetting (at least) in the stationary
limit $t\ra\infty$. That said, we shall henceforth focus our attention
on the model's \textit{stationary} density profiles.

We seek solutions for the stationary density profiles of the form
$\r_{\el}\asy\r\left(x\right)$ and $\t_{\el}\asy\t\left(x\right)$,
where $x=\el/L\in\co{0,1}$ denotes the macroscopic distance and $\asy$
is used to denoted asymptotic equality in the thermodynamic limit
$L\ra\infty$. If such solutions exist for all values of the problem's
parameters and satisfy the boundary conditions, they are the unique
solutions of the problem. We shall thus self-consistently assume this
scaling. Expanding out terms of the form 
\begin{equation}
u_{\el\pm1}=u\left(x\right)+\frac{1}{L}\partial_{x}u\left(x\right)+\frac{1}{2L^{2}}\partial_{x}^{2}u\left(x\right)+\ord{L^{-3}},\label{eq:expansion}
\end{equation}
while keeping the leading order behavior in large $L$, yields the
stationary bath density equation
\begin{equation}
0=\partial_{x}^{2}\r+\left(1-\frac{\e}{D}\right)\left(\r\partial_{x}^{2}\t-\t\partial_{x}^{2}\r\right),\label{eq:bath_den_simple-1-1}
\end{equation}
and the stationary tracer density equation
\begin{equation}
0=\partial_{x}^{2}\t-\left(1-\e\right)\left(\r\partial_{x}^{2}\t-\t\partial_{x}^{2}\r\right)-rL^{2}\left(1-\r_{0}-\t_{0}\right)\t.\label{eq:tracer_den_simple-1-1}
\end{equation}
It is interesting to notice that Eq. (\ref{eq:tracer_den_simple-1-1})
can only be affected by local resetting, diffusion, and exchange processes
if $\left(1-\r_{0}-\t_{0}\right)\pro L^{-2}$ as $L\ra\infty$, as
was previously noted and discussed in \cite{Miron2021}. Otherwise,
the only contribution to $\t$ would come from local resetting. For
now we keep the densities at the origin $\r_{0}$ and $\t_{0}$ as
parameters whose value will ultimately be set by demanding self-consistency
with the stationary MF profiles at $x=0$. This approach allows us
to restrict our analysis to the ``bulk'' of the system, i.e. sites
$\el\ne0$, explaining why the more general $\c_{\el}$ is replaced
by $\c_{\el\ne0}$ in the last term of Eq. (\ref{eq:tracer_den_simple-1-1}).

\subsection{Relating $\protect\r$ and $\protect\t$}

Our first step towards the solution of Eqs. (\ref{eq:bath_den_simple-1-1})
and (\ref{eq:tracer_den_simple-1-1}) involves relating the stationary
density profile $\r\left(x\right)$ to the stationary tracer density
profile $\t\left(x\right)$. Rewriting the bath density Eq. (\ref{eq:bath_den_simple-1-1})
as
\begin{equation}
-\r\left(1-\frac{\e}{D}\right)\partial_{x}^{2}\t=\left[1-\left(1-\frac{\e}{D}\right)\t\right]\partial_{x}^{2}\r,\label{eq:bath den relation 0}
\end{equation}
and performing integration-by-parts yields
\[
-\left(1-\frac{\e}{D}\right)\left[\r\left(y\right)\partial_{y}\t\right]\at{y=0}x+\left(1-\frac{\e}{D}\right)\int_{0}^{x}dy\partial_{y}\r\partial_{y}\t
\]
\begin{equation}
=\left[\left(1-\left(1-\frac{\e}{D}\right)\t\left(y\right)\right)\partial_{y}\r\right]\at{y=0}x+\left(1-\frac{\e}{D}\right)\int_{0}^{x}dy\partial_{y}\t\partial_{y}\r.\label{eq:bath den relation 1}
\end{equation}
The integrals in Eq. (\ref{eq:bath den relation 1}) cancel out, leaving
\begin{equation}
C_{0}-\left(1-\frac{\e}{D}\right)\r\partial_{x}\t=\left(1-\left(1-\frac{\e}{D}\right)\t\right)\partial_{x}\r,\label{eq:rho_tau_relation_1}
\end{equation}
where $C_{0}$ collects the boundary terms and is given by
\begin{equation}
C_{0}=\left(1-\frac{\e}{D}\right)\r\left(0\right)\partial_{y}\t\at{y=0}{}+\left(1-\left(1-\frac{\e}{D}\right)\t\left(0\right)\right)\partial_{y}\r\at{y=0}{}.\label{eq:C_0}
\end{equation}

We proceed by guessing that $C_{0}=0$ and will show a-posteriori
that this guess is correct. If $C_{0}=0$ Eq. (\ref{eq:rho_tau_relation_1})
becomes
\begin{equation}
-\frac{\left(1-\frac{\e}{D}\right)\partial_{x}\t}{1-\left(1-\frac{\e}{D}\right)\t}=\frac{\partial_{x}\r}{\r},\label{eq:bla}
\end{equation}
whose solution is
\begin{equation}
\r\left(x\right)=C_{1}\left(1-\left(1-\frac{\e}{D}\right)\t\left(x\right)\right),\label{eq:rho_tau_relation_2}
\end{equation}
where $C_{1}$ is an integration constant. To set its value we use
the conservation of the number of bath and tracer particles in Eq.
(\ref{eq:conservation_0}), transforming the discrete conservation
laws to integral form as 
\begin{equation}
\int_{0}^{1}dx\r\left(x\right)=\overline{\r}\text{ and }\int_{0}^{1}dx\t\left(x\right)=\overline{\t}.\label{eq:conservation}
\end{equation}
Integrating Eq. (\ref{eq:rho_tau_relation_2}) over the interval $\left[0,1\right]$
and using Eqs. (\ref{eq:conservation}) sets the integration constant
to
\begin{equation}
C_{1}=\frac{\overline{\r}}{1-\left(1-\frac{\e}{D}\right)\overline{\t}}.\label{eq:rho_tau_relation_3}
\end{equation}
We have thus derived the relation between the bath density profile
$\r\left(x\right)$ and the tracer density profile $\t\left(x\right)$
\begin{equation}
\r\left(x\right)=\overline{\r}\frac{1-\left(1-\frac{\e}{D}\right)\t\left(x\right)}{1-\left(1-\frac{\e}{D}\right)\overline{\t}}.\label{eq:rho_tau_relation_fin}
\end{equation}
Verifying self-consistency of the assumption $C_{0}=0$, for $C_{0}$
given in Eq. (\ref{eq:C_0}), is now immediate. Note that the denominator
in Eq. (\ref{eq:rho_tau_relation_fin}) only vanishes if $\frac{\e}{D}=1-\frac{1}{\overline{\t}}$,
which cannot happen since $0<\overline{\t}<1$ so the right-hand side
is always negative while both $\e$ and $D$ are positive. Another remark is that it is important to recall that Eq. (\ref{eq:rho_tau_relation_fin}) was derived under the assumption that the two stationary profiles are scaling functions of $x=\ell/L$. This assumption will next be shown to be self-consistent for local resetting, since $(1-\r_{0}-\t_{0})$ evidently scales as $\pro1/L^2$. 

\subsection{Tracer density profile}

We next use the relation between $\r\left(x\right)$ and $\t\left(x\right)$
in Eq. (\ref{eq:rho_tau_relation_fin}) to solve Eq. (\ref{eq:tracer_den_simple-1-1})
for $\t\left(x\right)$ which becomes
\begin{equation}
0=\partial_{x}^{2}\t-\a^{2}\t,\label{eq:tau_eqn}
\end{equation}
where the parameter $\a^{2}$ is defined as
\begin{equation}
\a^{2}\df L^{2}r\frac{1-\left(1-\frac{\e}{D}\right)\overline{\t}}{1-\left(1-\frac{\e}{D}\right)\overline{\t}-\left(1-\e\right)\overline{\r}}\left(1-\t_{0}-\r_{0}\right).\label{eq:alpha_sqr}
\end{equation}
The solution to Eq. (\ref{eq:tau_eqn}) is simply 
\begin{equation}
\t\left(x\right)=A_{-}e^{-\a x}+A_{+}e^{\a x},\label{eq:tau_sol_0}
\end{equation}
with the constants $A_{\pm}$ set using symmetry about the origin,
which dictates $\t\left(x\right)=\t\left(1-x\right)$, and tracer
number conservation $\int_{0}^{1}dx\t\left(x\right)=\overline{\t}$
of Eq. (\ref{eq:conservation}). The former yields $A_{-}=A_{+}e^{\a}$
and the latter $A_{+}=\frac{\a\overline{\t}e^{-\frac{\a}{2}}}{4\sinh\left[\frac{\a}{2}\right]}$,
so that we arrive at
\begin{equation}
\t\left(x\right)=\frac{\a\overline{\t}}{2}\frac{\cosh\left[\a\left(\frac{1}{2}-x\right)\right]}{\sinh\left[\frac{\a}{2}\right]}.\label{eq:tau_profile}
\end{equation}
The stationary tracer density profile $\t\left(x\right)$ has the
same form as that found in \cite{Miron2021}, which studied resetting
particles with exclusion. The only difference will appear when determining
the value of $\a$, which here also depends on the bath density at
the origin $\r_{0}$. The tracer density profile $\t\left(x\right)$
exhibits the characteristic tent-like shape, that was first obtained
for the stationary distribution of a single resetting Brownian particle
\cite{Evans_2011}. 

Since the bath density profile $\r\left(x\right)$ is related to the
tracer density profile $\t\left(x\right)$ via Eq. (\ref{eq:rho_tau_relation_fin}),
it may appear that our solution is complete. However, at this point,
$\a$ still remains an unknown parameter. An equation for $\a$ is
derived by demanding that $\t\left(x\right)$ at $x=0$ be consistent
with the tracer density at the origin
\begin{equation}
\t_{0}=1-\frac{\overline{\r}}{\frac{D}{\e}+\left(1-\frac{D}{\e}\right)\left(\overline{\r}+\overline{\t}\right)}+\ord{L^{-2}}.\label{eq:tau_0}
\end{equation}
Equation (\ref{eq:tau_0}) is obtained by first using Eq. (\ref{eq:rho_tau_relation_fin})
to get $\r_{0}=\overline{\r}\frac{1-\left(1-\frac{\e}{D}\right)\t_{0}}{1-\left(1-\frac{\e}{D}\right)\overline{\t}}$,
then solving Eq. (\ref{eq:alpha_sqr}) for $\t_{0}$ to leading order
in $L$, and finally demanding agreement with $\t\left(x\right)$
of Eq. (\ref{eq:tau_profile}) at $x=0$. The result is a transcendental
equation for $\a$ 
\begin{equation}
\frac{\a\overline{\t}}{2\tanh\left[\frac{\a}{2}\right]}=1-\frac{\overline{\r}}{\frac{D}{\e}+\left(1-\frac{D}{\e}\right)\left(\overline{\r}+\overline{\t}\right)},\label{eq:alpha eqn}
\end{equation}
which we must solve numerically to compute the profiles. The denominator
on the right-hand side of Eq. (\ref{eq:alpha eqn}) vanishes if $\frac{\e}{D}=1-\frac{1}{\overline{\r}+\overline{\t}}$
but, since $\frac{\e}{D}>0$ and $0<\overline{\r}+\overline{\t}<1$,
this equality cannot be satisfied as $1-\frac{1}{\overline{\r}+\overline{\t}}$
is negative.

\begin{figure}[H]
\begin{centering}
\includegraphics[scale=0.75]{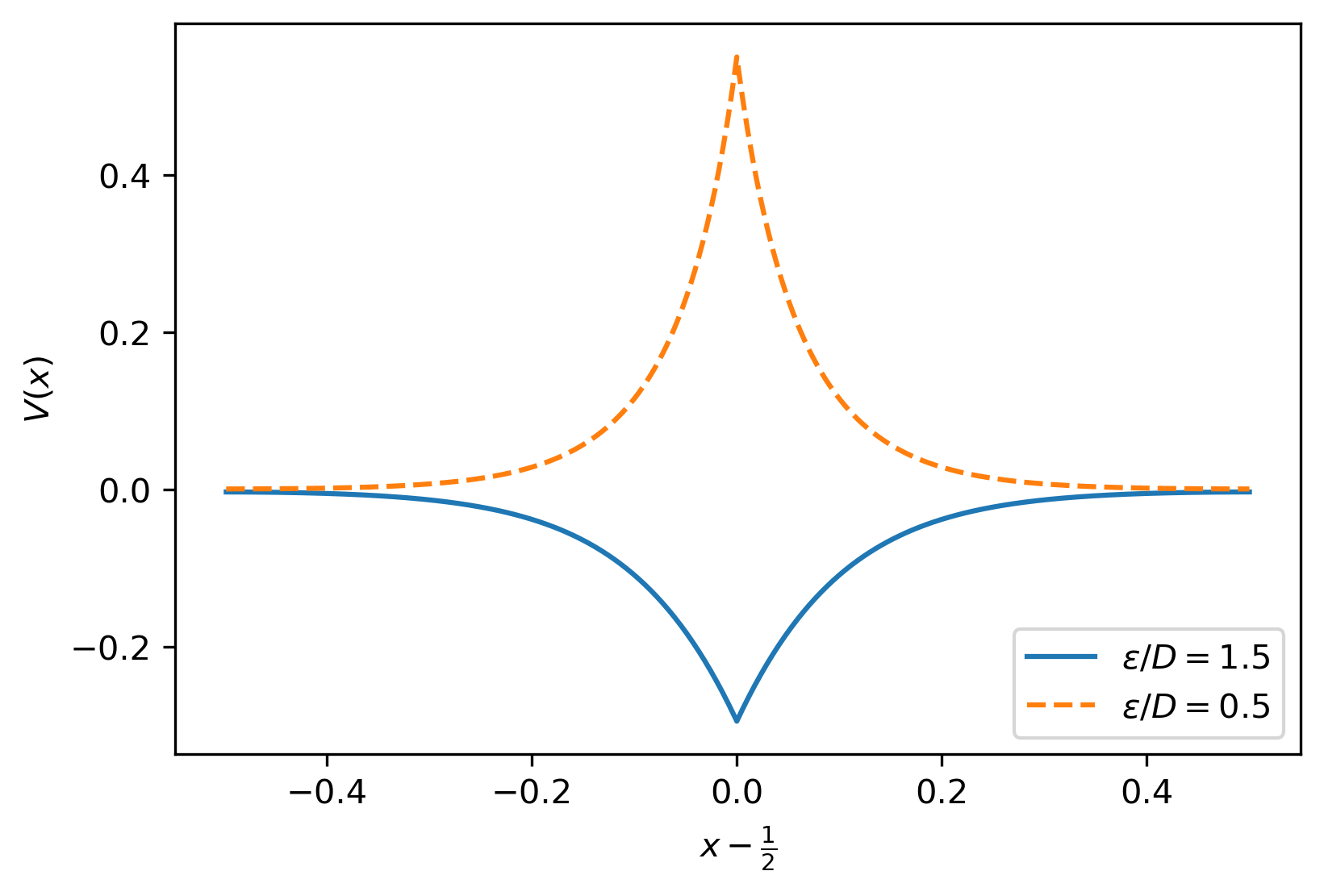}
\par\end{centering}
\caption{Comparison of the MF effective potential $V\left(x\right)$ appearing
in Eq. (\ref{eq:V(x)}) for $\protect\e/D=1.5$ ($\protect\e=1.5$
and $D=1$) and for $\protect\e/D=0.5$ ($\protect\e=1$ and $D=2$),
with model parameters $\overline{\protect\r}=1/4$, $\overline{\protect\t}=1/8$,
and $r=1$. These values of $\protect\e$ and $D$ correspond to those
considered in Fig. \ref{differences eps/D !=00003D 1}, showing the
resetting tracers behave as an attractive potential for $\protect\e/D=1.5$
and a repulsive potential for $\protect\e/D=0.5$.}

\label{V(x)}
\end{figure}

The set of Eqs. (\ref{eq:rho_tau_relation_fin}), (\ref{eq:tau_profile}),
and (\ref{eq:alpha eqn}) constitute the stationary MF solutions to
the bath and tracer density profiles. An inspection of Eq. (\ref{eq:rho_tau_relation_fin})
reveals that the ratio $\frac{\e}{D}$, which describes the competition
between the degree of geometric constraints and the bath diffusivity,
plays a crucial role in shaping the bath density profile $\r\left(x\right)$.
Specifically, it determines the sign of $\r\left(x\right)$'s curvature.
Taking a second derivative of the relation between $\r\left(x\right)$
and $\t\left(x\right)$ in Eq. (\ref{eq:rho_tau_relation_fin}) gives
\begin{equation}
\frac{\partial_{x}^{2}\r}{\partial_{x}^{2}\t}=\frac{\overline{\r}\left(\frac{\e}{D}-1\right)}{1-\left(1-\frac{\e}{D}\right)\overline{\t}}.\label{eq:curvature}
\end{equation}
Local resetting causes the tracer density to be highest at the origin,
while the tracers' exchange and diffusive dynamics ensure that it
falls off rapidly (faster than linear) with distance from the origin.
This leads to a positive curvature for $\t\left(x\right)$. As such,
the curvature of $\r\left(x\right)$ is set by the right-hand side
of Eq. (\ref{eq:curvature}). We immediately notice the two limits
$\frac{\e}{D}\ra0$ and $\frac{\e}{D}\ra\infty$, respectively corresponding
to $\partial_{x}^{2}\r=-\frac{\overline{\r}}{1-\overline{\t}}\partial_{x}^{2}\t<0$
and $\partial_{x}^{2}\r=\frac{\overline{\r}}{\overline{\t}}\partial_{x}^{2}\t>0$.
A closer look shows that the right-hand side of Eq. (\ref{eq:curvature})
is negative for $\frac{\e}{D}<1$ and positive for $\frac{\e}{D}>1$,
with $\r\left(x\right)$'s curvature changing sign at $\frac{\e}{D}=1$.
This result can be understood as follows: When $D>\e$ the bath particles
manage to diffuse away from the high tracer-density region surrounding
the origin. Since the high-density region decays strongly, once outside
its reach bath particles quickly spread out and inhabit the remaining
regions, implying a negative curvature for $\r\left(x\right)$. But
when $D<\e$ the bath particles are too slow and become trapped inside
the high-density region. The bath density thus remains slightly higher
near the origin and slowly falls-off as the tracer density diminishes,
indicative of a positive curvature for $\r\left(x\right)$. The transition
in $\r\left(x\right)$'s curvature is also reflected in the effective
MF potential $V\left(x\right)$ that the bath particles experience
due to the tracer profile (see Fig. \ref{V(x)}). To find $V\left(x\right)$
we compare $\r\left(x\right)$ to the equilibrium Boltzmann distribution
$\overline{\r}\frac{e^{-\b V\left(x\right)}}{Z}$, where $Z=\int_{0}^{1}dxe^{-\b V\left(x\right)}$
is the partition function. We find
\begin{equation}
Z=\left(1-\left(1-\frac{\e}{D}\right)\overline{\t}\right)^{-1},\label{eq:Z}
\end{equation}
and the potential $V\left(x\right)$ 
\begin{equation}
V\left(x\right)=-\frac{\log\left[1-\left(1-\frac{\e}{D}\right)\t\left(x\right)\right]}{\b}.\label{eq:V(x)}
\end{equation}
Notice that the bath particles feel no external potential when $\e/D=1$,
which is consistent with their homogeneous profile. For $\e/D\ne1$
we can find the shape of $V\left(x\right)$ near the origin by substituting
$\t\left(x\right)$ of Eq. (\ref{eq:tau_profile}) into $V\left(x\right)$
and expanding around $x=0$. This gives 
\begin{equation}
V\left(x\right)=V_{0}-\frac{\a^{2}\overline{\t}\left(1-\frac{\e}{D}\right)}{\a\overline{\t}\left(1-\frac{\e}{D}\right)\coth\left[\frac{\a}{2}\right]-2}\left|x\right|+\ord{x^{2}},\label{eq:V(x) expansion}
\end{equation}
with $V_{0}=\log\left[1-\frac{\a}{2}\coth\left[\frac{\a}{2}\right]\left(1-\frac{\e}{D}\right)\overline{\t}\right]$,
which is a linear potential whose slope can be shown to change sign
at $\frac{\e}{D}=1$.

We conclude this section with two additional points that are worth
mentioning. As found in \cite{Miron2021}, here too the resetting
rate $r$ never enters the profiles in the limit $L\ra\infty$. This
can be heuristically understood from the fact that, in this limit,
there is an infinite number of tracers in the system. Correspondingly,
there is an infinite number of tracers attempting to reset their positions
to the origin at any given moment. As long as the resetting attempt
rate $r$ is finite (and independent of $L$), its precise value is
irrelevant. Another interesting point is that, although Eqs. (\ref{eq:bath_den_simple-1-1})
and (\ref{eq:tracer_den_simple-1-1}) depend on $\e$ and $D$ separately,
the MF profiles depend only on the ratio $\frac{\e}{D}$ in the thermodynamic
limit. It can be shown that $r$, $\e$, and $D$ do individually
enter the higher order, finite-size corrections to the asymptotic
profiles.

\section{Conclusion and discussion \label{sec:Conclusion-and-discussion}}

This work studies the interplay of geometric constraints and ``local
resetting'', where particles attempt stochastic resetting independently
of one-another, thus introducing interactions into the resetting process.
The dynamics associated with the classical setup of diffusive spherical
particles with short-ranged repulsive interactions confined to a narrow
channel is extended to include local resetting, and then its main
ingredients are encoded into the dynamics of hopping particles on
a $1D$ lattice ring. The lattice contains two particles species,
locally resetting ``tracers'' of mean density $\overline{\t}$ and
non-resetting ``bath'' particles of mean density $\overline{\r}$.
Their continuous time evolution dynamics mimic those of the narrow
channel setup: hard-core exclusion replaces the short-ranged interactions,
the different masses are modeled by setting the bath and tracer diffusion
rates to $1$ and $D$ respectively, and the channel's width is replaced
by a finite exchange rate $\e$. Evolution equations for the mean
bath and tracer density profiles are derived, and then solved in the
stationary limit using the mean-field (MF) approximation in the thermodynamic
limit $L\ra\infty$. The MF tracer density profile exhibits the typical
tent-like shape with a cusp at the origin, as found for a single resetting
diffusive particle \cite{Evans_2011}. Yet the interplay of local
resetting and geometric confinement manifests most dramatically in
the MF bath density profile $\r\left(x\right)$, which transitions
between ``repelled'' and ``trapped'' states as the value of $\e/D$
crosses $1$. For $\e/D<1$ bath particles are strongly repelled from
the origin, with $\r\left(x\right)$ taking the shape of an inverted
tent with a negative curvature $\r''\left(x\right)<0$. For $\e/D>1$
the bath particle are too slow to escape the dense origin and remain
trapped. The shape of $\r\left(x\right)$ is then similar to that
of $\t\left(x\right)$ and $\r''\left(x\right)>0$. For $\e/D=1$
the bath particles experience a homogeneous environment, since hopping
into a vacant site and exchanging with a tracer happen with the same
rate, and we obtain $\r\left(x\right)=\overline{\r}$ . While the
MF approximation successfully predicts the existence of both states
and the transition between them, there is no a-priori reason for it
to be exact in any regime. However, numerical investigations of the
model provide strong evidence to suggest that the MF approximation
does in fact become exact for $\e=D=1$ as $L\ra\infty$. This result,
whose underlying origins remain unclear, joins the observations of
\cite{Miron2021,Pelizzola_2021} where the MF approximation was found
to stand in unexpectedly good agreement with numerical simulation
results of models exhibiting local resetting. 

Looking forward, many interesting open questions still remain unsolved
and demand attention. First and foremost, why does the MF approximation
seem to work well in models featuring local resetting? Is there any
way to adapt exact methods like the matrix product ansatz \cite{blythe2007nonequilibrium,kriecherbauer2010pedestrian},
which have been used to derive exact solutions for various lattice
models with exclusion interactions, to treat local resetting? Another
exciting direction is to explore the temporal evolution in the presence
of local resetting. While this is interesting for local resetting
in general, as dynamical phase transitions were found in the temporal
evolution of the density profile of a single resetting Brownian particle
\cite{evans2020stochastic}, it is even more interesting in the context
of geometric confinement and tracer sub-diffusion \cite{jepsen1965dynamics,percus1974anomalous,alexander1978diffusion}.
In addition, both experimental and theoretical research following
this line of work would of greatly benefit from pursuing a quantitative
relation between the parameters describing the narrow channel setup,
and the corresponding parameters of the lattice model.

\section{Acknowledgments}

I wish to thank David Mukamel, Oren Raz, and Shlomi Reuveni for critically
reading this manuscript and for their very helpful suggestions and
remarks. But, most crucially, hhis work has been made possible by the grace of DBM and
the mercy of GM, to whom I am indebted to for their unconditional
support.

\end{document}